\documentclass[twocolumn, twocolappendix]{aastex631}
\usepackage{amsmath}

\newcommand{\Height}{0.42}
\newcommand{\Width}{0.18}
\newcommand{\Heighta}{0.44}
\newcommand{\Widtha}{0.2}
\begin{document}

\title{Flux Density Stability and Temporal Changes in the Spectra of Millisecond Pulsars using the GMRT}
\correspondingauthor{Rahul Sharan}
\email{rsharan.ncra@gmail.com}

\author[0009-0001-9428-6235]{Rahul Sharan}
\affiliation{National Centre for Radio Astrophysics, NCRA-TIFR, Pune 411007, India}

\author[0000-0002-6287-6900]{Bhaswati Bhattacharyya}
\affiliation{National Centre for Radio Astrophysics, NCRA-TIFR, Pune 411007, India}

\author[0000-0002-3764-9204]{Sangita Kumari}
\affiliation{National Centre for Radio Astrophysics, NCRA-TIFR, Pune 411007, India}

\author[0000-0002-2892-8025]{Jayanta Roy}
\affiliation{National Centre for Radio Astrophysics, NCRA-TIFR, Pune 411007, India}

\author[0009-0002-3211-4865]{Ankita Ghosh}
\affiliation{National Centre for Radio Astrophysics, NCRA-TIFR, Pune 411007, India}
 
\begin{abstract}
This paper presents an investigation of spectral properties of 10 millisecond pulsars (MSPs) discovered by the uGMRT, observed from 2017--2023 using band 3 (300--500 MHz) and 4 (550--750 MHz) of uGMRT. For these MSPs, we have reported a range of spectral indices from $\sim$0 to $-$4.8, while averaging the full observing band and all the observing epochs. For every MSP, we calculated the mean flux densities across 7--8 sub-bands each with approximately 25 MHz bandwidth spanning band 3 and band 4. We computed their modulation indices as well as average and maximum-to-median flux densities within each subband. Using a temporal variation of flux density we calculated the refractive scintillation time scales and estimated structure function with time lag for 8 MSPs in the sample. We note a significant temporal evolution of the in-band spectra, classified into three categories based on the nature of the best-fit power-law spectra, having single positive spectral indices, multiple broken power law, and single negative spectral indices. Additionally, indications of low-frequency turnover and a temporal variation of the turnover frequency (to the extent that turnover was observed for some of the epochs while not seen for the rest) were noted for all the MSPs. To the best of our knowledge, this is the first systematic investigation probing temporal changes in the MSP spectra as well as in turnover frequency. Future exploration with dense monitoring combined with modeling of spectra can provide vital insight into the intrinsic emission properties of the MSPs and ISM properties.
\end{abstract}


\section{Introduction} \label{sec:intro}
Understanding the spectra of pulsars is crucial for deciphering the pulsar emission mechanism and deriving the pulsar luminosity function across a broad spectrum of radio frequencies \citep{Gupta_et_al_2003, Philippov_Krammer_2022}.  
For quite some time, it has been established, as evidenced by \cite{Sieber_1973}, that radio pulsars exhibit a relationship between flux density (S$_\nu$) and frequency ($\nu$), wherein the observed flux density experiences a pronounced decline as frequency increases. This pattern is frequently represented by a power-law given by, $S_{\nu}=\nu^{\alpha}$, where $\alpha$ is the spectral index. \cite{Lorimer_et_al_1995} conducted an analysis of spectral indices for a substantial number of normal pulsars (with spin period $>$30 ms), presenting spectra for 280 pulsars derived from flux density measurements obtained across up to five distinct radio frequencies ranging from 0.4 to 1.6 GHz, utilizing the Lovell radio telescope and other estimates available in the literature. This study concluded that the spectra can be described by a simple power law, over the frequency range considered. 

Combining flux density estimates available in the literature, \cite{Lorimer_et_al_1995} estimated a mean spectral index $\alpha$ of $-$1.6. More recent studies (e.g. \cite{Jankowski_et_al_2018} for a frequency range 728$-$3100 MHz including the flux density estimates available in the literature) found that the average $\alpha$ for the normal pulsars (spin period $>$30 ms) ranges between $-$1.1 to $-$2.1, whereas \cite{Han_et_al_2016} found the distribution of $\alpha$ to vary from $-$0.4 to $-$4.8, using Parkes multi-beam receiver operating at central frequency of 1369 MHz with a bandwidth of 256 MHz. \cite{Bates_et_al_2013} determined a mean $\alpha$ of $–$1.4 with a unit standard deviation, consistent with a Gaussian distribution. Recent studies also found that, for a subset of pulsars with multi-frequency observations, a single power-law model does not adequately match the observed data \citep{Jankowski_et_al_2018}, and two (or more) power-law models 
(typically one for low frequency and the other for higher frequency) are required. Some pulsars have exhibited spectral turnover at low frequencies, under 400 MHz, \citep[e.g. ][]{Sieber_1973, Kuzmin_et_al_1978, Izvekova_et_al_1981, Malofeev_et_al_1994}, while a smaller sample has shown similar characteristics at relatively higher frequencies close to 1 GHz \citep{Maron_et_al_2000, Kijak_et_al_2011}. We note that the majority of the investigations of pulsar spectra considered non-simultaneous observations at different frequencies as well as an average value of the flux density is quoted for the full observing bandwidth of the telescopes. While averaging over the full observing band could provide a good approximation of spectra for a narrow-band telescopes, for the present and upcoming wide bandwidth telescope this will lead to erroneous estimates. 

Even though the millisecond pulsar (MSP; having spin period $<$30 ms) population has drastically increased in the last two decades (by a factor of 4\footnote{\url{https://www.astro.umd.edu/~eferrara/pulsars/GalacticMSPs.txt}}), investigation of MSP spectra is still in its infancy as compared to normal
pulsars. The reason for this is that MSPs are inherently faint sources, coupled with the limited access to highly sensitive wideband observing tools until approximately the last decade (prior to the enhanced capabilities of upgraded GMRT \citep{uGMRT_ref_2017}, the UWL receiver \citep{Parkes_UWL} for Parkes, and the activation of LOFAR \citep{Lofar_2013}). No significant changes in the spectral properties between the normal pulsars and MSPs are reported by \cite{Kramer_et_al_1998}, and \cite{Dai_et_al_2015}. On the other hand, investigations by \cite{Foster_et_al_1991}, and \cite{Toscano_et_al_1998} reported a relatively steeper spectral nature of MSPs compared to normal pulsars. 

\cite{Gitika_et_al_2023} shows a single power law can reproduce the spectra for most of the MSPs in their sample observed in the L-band using the MeerKAT receiver (856--1712 MHz, \cite{Meerkat_2016}). \cite{Kondratiev_et_al_2016} and \cite{Kuzmin_et_al_2002} shows no indication of low frequency turnover. On the other hand, \cite{Wang_et_al_2021} noticed evidence of turnovers at around 350 MHz for J0318$+$0253 (an MSP), and \cite{Kuniyoshi_et_Al_2015} saw turnover at frequencies as low as 100 MHz in 10 out 39 sample MSPs. Some other studies \citep{Kramer_et_al_1998, Dai_et_al_2015, Lorimer_et_al_1995} also noticed that single power law is not the best fitting curve for spectra for few MSPs in their sample. 

The flux density measurements for pulsars show the signature of refractive scintillation on a timescale from weeks to months \citep{Rickett_et_al_1984, Stinebring_et_al_1990, Sieber_1973, Gitika_et_al_2023}. Electron-density irregularities on a larger scale (10$^{11}$—10$^{13}$ cm) can lead to the focusing and defocusing of rays originating from pulsars (\cite{Rickett_et_al_1984}).  Refractive scintillation is believed to be caused by the large-scale structures in the strong scattering regime \citep{Prokhorov_et_al_1975, Rickett_et_al_1984, LorimerKramer}. These large spatial scales have the potential to generate gradual intensity fluctuations over extended periods (ranging from days to years) and give rise to refractive interstellar scintillations (RISS) in Galactic pulsars. The flux density variations on the time scale of weeks to months are often associated with refractive scintillation. There are other reasons for such variations like intrinsic flux density variations, neighboring environments \citep{Rajwade_et_al_2015, Kijak_et_al_2011}, in the case of binary MSPs, the possible effect of companion (e.g. change of spectra just before and after eclipsing phases \cite{Kansabanik_et_al_2021}), and other effects of ISM like extreme scattering events \citep{Zhu_et_al_2023, Fiedler_et_al_1994}, plasma lensing \citep{Lin_et_al_2023}.

In this paper, we present the result from flux density measurements and temporal variation of the spectral properties for 10 MSPs discovered by the uGMRT observed for 5 years. With the recent availability of telescopes like uGMRT, UWL receiver of Parkes, and MeerKAT, enabling us to probe the flux density variation with the wide observing bandwidth, we need to consider the flux density variation for the observing band divided into multiple subbands (will be referred to as in-band spectra in the rest of the paper and is detailed in Section \ref{sec:in-band_spectra}). Section \ref{sec:Obs} describes the observations presented in this paper. The offline data analysis procedure is detailed in Section \ref{sec:offline_data_analysis}. In Section \ref{results} results from this study are presented, where Section \ref{sec:Temp_variation_of_tot_flux} reports temporal variation of flux density, Section \ref{sec:Model_structure_func} presents an analysis of refractive scintillation parameters and Section \ref{subsec:Temp_variation_of_in_band_spectra} deals with temporal variations of the in band spectra. Finally, Section \ref{discussion} presents a discussion on the above results.

\section{Necessity of In-band spectra}
\label{sec:in-band_spectra}
\begin{figure}[!hbt]
\hspace*{-1.1cm}
\includegraphics[scale=0.4]{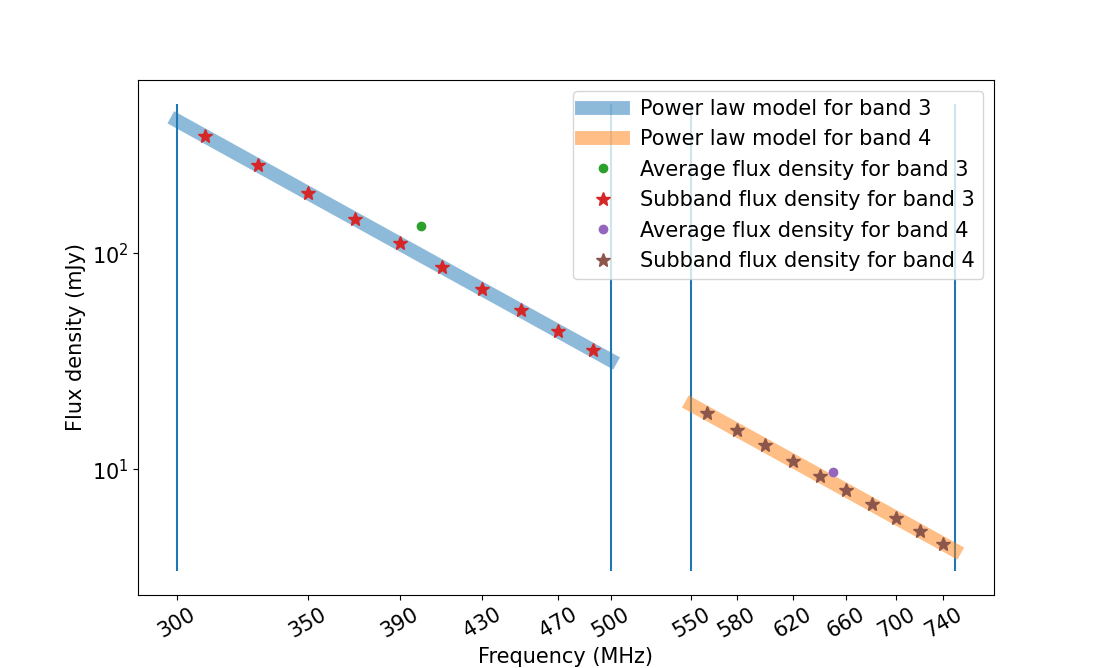}
\hspace*{1cm}
\caption{This plot illustrates the deviation of the average flux density points for subband versus full-band averaged spectral index calculation for uGMRT observations across band 3(300--500 MHz) and band 4(550--750 MHz). The blue and orange lines (which are thick but faded lines) represent the spectral index of $-5$ in band 3 and band 4 respectively. The lines with connected red and brown stars respectively in band 3 and band 4 present the flux calculated by taking 9 subbands of the full wideband frequency range. The green and violet points represent the full-band averaged flux density for band 3 and band 4 respectively. The resultant spectral index $\alpha$ using the full wideband averaged flux density comes out to be $-5.411$. This is due to the central frequency dependent correction terms as described in Section \ref{sec:in-band_spectra}
\label{fig1:spectra_correction_a_5}}
\end{figure}

The observed flux density is obtained by averaging over a range of frequencies (say between $\nu_{1}$ and $\nu_{2}$). Due to this, as seen in Equation \ref{eq1:average_flux}, the observed flux density has an additional factor, $f_{\alpha}(\nu_{1}, \nu_{2})$, due to averaging. This factor, $f_{\alpha}(\nu_{1}, \nu_{2})$, is nearly equal to unity for wideband observations. But on the contrary, wideband telescopes like uGMRT, Parkes with UWL receivers, etc, where bandwidth is comparable to the central observing frequency, $f_{\alpha}(\nu_{1}, \nu_{2})$ plays a key role while calculating the full-band averaged flux density. The average flux density ($\langle S_{\nu} \rangle$) is described by the following equation,
\begin{subequations}\label{eq1:flux_averaging}
\begin{align}
\langle S_{\nu} \rangle &= \frac{\int_{\nu_1}^{\nu_2} S_{\nu} d{\nu}}{\int_{\nu_1}^{\nu_2} d\nu}\label{eq1:average_flux} \\
 & = S_{\bar{\nu}}f(\nu_1, \nu_2)
\end{align}

In astrophysics, the power law is often the preferred model for spectral flux density, expressed as $ S_{\nu} = S_0\nu^{\alpha} $, where $S_0$ represents the flux at a reference frequency $\nu_0$.
\begin{equation}
f_{\alpha}(x) = \frac{\left(1 + \frac{x}{2}\right)^{\alpha + 1} - \left(1 - \frac{x}{2}\right)^{\alpha + 1}}{\left( \alpha + 1 \right)x}\label{eq1:averaging_for_spectrae}
\end{equation}
\end{subequations}
where $x = \frac{\Delta\nu}{\Bar{\nu}}$ ; $ \Delta\nu$, bandwidth ($ = \nu_2 - \nu_1 $)and $\Bar{\nu}$, central frequency ($ = \frac{\nu_2 + \nu_1 }{2}$).
The factor, $f_{\alpha}(x)$ $\sim$ 1 for $x \ll 1$, which is true for narrowband observations. Whereas for wideband observations, $x$ is of the order of unity, thus $f_{\alpha}(x) > 1$. Hence there will be a difference in the calculated and intrinsic flux density. Table \ref{Table1: flux_averaging} presents the computed values of $f_{\alpha}(x)$ for some of the wideband telescopes (e.g. uGMRT, Parkes, and LOFAR), considering $\alpha$ as $-2$ (typical value of the spectral index) and $-5$ (extreme values of the spectral index). The change in values of $f_{\alpha}(x)$ with different observing bands are more severe for wideband telescopes than that for narrowband observations. Such high values of the correction factor $f_{\alpha}(x)$, can be solved by sub-banding the full observing frequency band. This introduces an additional frequency (i.e. central frequency of the band) dependent term and is described in Figure \ref{fig1:spectra_correction_a_5}.

\begin{table*}[hbt!] 
\caption{Additional averaging factor, $f_{\alpha}(x) $, on the calculated flux density for various wideband telescopes}
\begin{center}

\vspace{0.2cm}
\begin{tabular}{|l|c|c|c|c|c|}
\hline

Telescope Name & Bandwidth (MHz) & Central frequency (MHz)  & $f_{\alpha}(x) $ at $\alpha = -5$ & $f_{\alpha}(x) $ at $\alpha = -2$ \\\hline
uGMRT (band 3)    &  200 & 400  & 1.375 & 1.066  \\\hline
uGMRT (band 4)    &  200 & 650  & 1.126 & 1.024 \\\hline
uGMRT (band 5)   &  400 & 1260  & 1.135 & 1.025 \\\hline
Parkes (UWL reciever)    &  3328 & 2368  & 22.759 & 1.927 \\\hline
\href{https://science.astron.nl/telescopes/lofar/lofar-system-overview/technical-specification/frequency-subband-selection-and-rfi/}{LOFAR}     &  80 & 50  & 97.118 & 2.778\\\hline
\href{https://www.skao.int/en/explore/telescopes/ska-low}{SKA-Low} & 300 & 200 & 42.648 & 2.285 \\\hline

\end{tabular}
\end{center}
Although the band dependent factors are relatively small for uGMRT, the calculation of spectral index using full-band averaged flux density for band 3 and band 4 of uGMRT generates a significant error (as demonstrated in Figure \ref{fig1:spectra_correction_a_5}) from the assumed model. We truncated the decimal values up to the third decimal place as the $f_{\alpha}(x)$ for band 4 and band 5 shows its effect from the third decimal place (for $\alpha=-2$). This change in the correction factor might seem small at first glance but its effect is seen in the estimated parameter $\alpha$.
\label{Table1: flux_averaging}
\end{table*}


    
\section{Observations} \label{sec:Obs}
The spectral properties of MSPs presented in this paper are the outcome of a series of follow-up observations carried out for the target MSPs discovered by the GMRT, both from GHRSS, \cite{GHRSS_1}, \cite{Bhattacharyya_et_al_2019} and FERMI-directed searches, \cite{FERFI_LAT_1}, \cite{Surrendipitous_discovery}.  
These observations were performed using, uGMRT \citep{uGMRT_ref_2017}, an array of 30 antennas (each having a diameter of 45 m), operating over 4 bands between 100--1460 MHz. The GMRT is a highly sensitive and wide-bandwidth instrument for investigating MSP in-band spectra with sub-banding the observing band.  

Regular follow-up observations of the target MSPs were performed between 2017 to 2023 in the phased array (PA) with coherent dispersed (CD) or incoherent dedispersed (ID) observing mode of uGMRT in band 3 (300--500 MHz), and band 4 (550--750 MHz). The last column of Table \ref{Table2: MSP_list}, shows the summary of observations of the MSPs. In our sample, seven MSPs are in binary systems and three of these MSPs, namely J1242$-$4712, J1544$-$4937, and J2101$-$4802, are spider MSPs \citep{Roberts_2013} in compact binary orbits. We also observed three isolated MSPs namely J0248$+$4230, J1207$-$5050, and J1646$-$2142.
Observations were taken with 4096 frequency channels with a time resolution of 81.96 $\mu$s for ID mode and 512 channels with a time resolution of 20 $\mu$s for CD mode. 


\begin{table*}[!htb]
\caption{List of parameters and observational specifics for the target MSPs.}
\begin{center}

\vspace{0.2cm}
\begin{tabular}{|l|c|c|c|c|c|c|}
\hline

Pulsar Name & Period (ms) & DM (pc/cm$^3$) & \multicolumn{1}{|p{2cm}|}{\centering Scattering time$^{\dagger}$ \\ at 300 MHz (in $\mu$s)} & \multicolumn{1}{|p{2cm}|}{\centering Decorrelation bandwidth (in kHz) }& Total number of epochs (B3, B4, B5)\\\hline
J0248+4230 & 2.6 & 48.263 & 70.87 & 2.245 & 30 (28/31, 2/3, 0/0) \\\hline
J1120-3618 & 5.55 & 45.125 & 55.18 & 2.884 & 41 (37/37,3/4,0/1) \\\hline
J1207-5050 & 4.84 & 50.673 & 85.21 & 1.867 & 28 (0/1, 20/20, 8/9) \\\hline
J1242-4712 & 5.31 & 78.647 & 495.73 & 0.321 & 15 (10/11, 4/5, 1/1) \\\hline
J1536-4948 & 3.07 & 38.001 & 29.65 & 5.36 & 36 (18/18, 8/9, 0/2) \\\hline
J1544+4937 & 2.14 & 23.224 & 5.82 & 27.33 & 63 (37/37, 26/26, 0/0) \\\hline
J1646-2142 & 5.85 & 29.741 & 12.82 & 12.413 & 49 (30/32, 17/17, 2/5) \\\hline
J1828+0625 & 3.62 & 22.428 & 5.23 & 30.417 & 27 (29/30, 8/9, 0/0) \\\hline
J2101-4802 & 9.48 & 25.055 & 7.37 & 21.586 & 16 (16/16, 0/0, 0/0) \\\hline
J2144-5237 & 5.04 & 19.546 & 3.47 & 45.832 & 52 (29/34, 9/9, 10/10) \\\hline

\end{tabular}
\end{center}
$^{\dagger}$ Calculated using emperical formula from \cite{Bhatt_et_al_2004}.

In the last column, the number outside the bracket is the total number of observing epochs for each MSP, the numerator is the  number of epochs for detection, and the denominator is the total number of the observing epochs at each band (3,4,5). The typical observing time per epoch ranges from 30 min to 2 hrs.

\label{Table2: MSP_list}
\end{table*}

\begin{table*}
\caption{Table containing average flux density S$_{mean}$, modulation index m$_{ind}$, parameter R, average spectral index ($\alpha$) using band 3 (B3) and 4 (B4), refractive timescales $\tau_r$ and the slope of structure function ($\alpha_0$ ) with a time lag in logarithmic space at the structure regime.}\label{Table3: PSR_summary_info}
\begin{center}
\hspace*{-4cm}
\begin{tabular}{|l|c|c|c|c|c|c|c|c|c|c|c|}

\hline
PSR name &B$3$ S$_{mean}$ &B$3$ m$_{ind}$ &B$3$ R &B$4$ S$_{mean}$ &B$4$ m$_{ind}$ &B$4$ R & \multicolumn{1}{|p{1.5cm}|}{\centering  DM \\ (pc/cm$^3$)} &\multicolumn{1}{|p{2cm}|}{\centering Spectral Index ($\alpha$) \\ using \\ B3 and B4} & \multicolumn{1}{|p{2cm}|}{\centering Refractive Timescale \\ ($\tau_r$) in days } & $\alpha_0$ \\\hline
J0248+4230$^{\gamma}$ & 0.77(1) & 0.453 & 2.989 & 0.26(2) & - & - & 48.27 & -2.163 & $5878.5 \pm 1391.31$ & $3.28 \pm 2.77$ \\\hline
J1120-3618 & 0.99(2) & 0.775 & 4.898 & 0.19(1) &  -  & - & 45.125 & -3.37 & $40.25 \pm 20.5$ & $1.81 \pm 0.24$\\\hline
J1207-5050$^{\gamma}$ & - & - & - & 0.94(2) & 0.714 & 3.258 & 50.673 & - & - & - \\\hline
J1242-4712$^{\beta}$ & 0.76(2) & 0.539 & 2.343 & 0.75(3) & - & - & 78.648 & -0.006 & - & - \\\hline
J1536-4948 & 10.7(3) & 0.848 & 4.99 & 1.02(4) & 0.879 & 4.238 & 38.001 & -4.845 & $29.34 \pm 3.06$ & $3.31 \pm 6.01$ \\\hline
J1544+4937$^{\beta}$ & 1.96(3) & 0.509 & 2.274 & 0.78(1) & 0.373 & 1.748 & 23.224 & -1.888 & $>1000$ & $NA$\\\hline
J1646-2142$^{\gamma}$ & 1.47(3) & 0.818 & 5.848 & 0.79(2) & 0.51 & 2.968 & 29.741 & -1.266 & $>1000$ & $NA$\\\hline
J1828+0625 & 0.78(1) & 0.817 & 4.873 & 0.31(1) & 0.287 & 1.568 & 22.418 & -1.884 & $1737.87\pm 83.25$ & $6.1\pm 5.68$ \\\hline
J2101-4802$^{\beta}$ & 1.63(5) & 0.555 & 1.752 & - & - & - & 25.055 & - & - & - \\\hline
J2144-5237 & 1.21(3) & 0.994 & 6.476 & 0.82(3) & 0.903 & 4.444 & 19.546 & -0.796 & $67.84 \pm 15.61$ & $2.69 \pm 1.02$ \\\hline
\end{tabular}
\hspace*{1cm}

\end{center}
$\gamma$: Isolated MSP, $\beta$: Spider MSP; 
Note: For a given band modulation index and parameter $R$ with MSPs less than 8 epochs of observation aren't calculated. 
\end{table*}
\section{Offline Data Analysis} \label{sec:offline_data_analysis}
\subsection{Flux density calculation}
\label{sec:pipeline_details}
The uGMRT data is recorded in filterbank format which is then run through the \textsc{GPTOOL}\footnote{\url{https://github.com/chowdhuryaditya/gptool}} to remove the RFI by identifying the outlier and correct for the bandshape of the data. This RFI mitigated and bandshape corrected filterbank file is then folded using the parameters of the pulsars, derived from the timing studies \citep{Surrendipitous_discovery, Bhattacharyya2022, GHRSS_1, Kumari_et_al_2023, Ghosh_et_al_2024, Sharma_et_al_2024} and stored in the folded data cube format (power, i.e. stokes I, as a function of time(split in sub integration), frequency(split in subband) and phase of the pulse) using \textsc{PRESTO} \citep{PRESTO_ref}.

\begin{figure}[!hbt]
\hspace*{-1.1cm}
\includegraphics[scale=0.24]{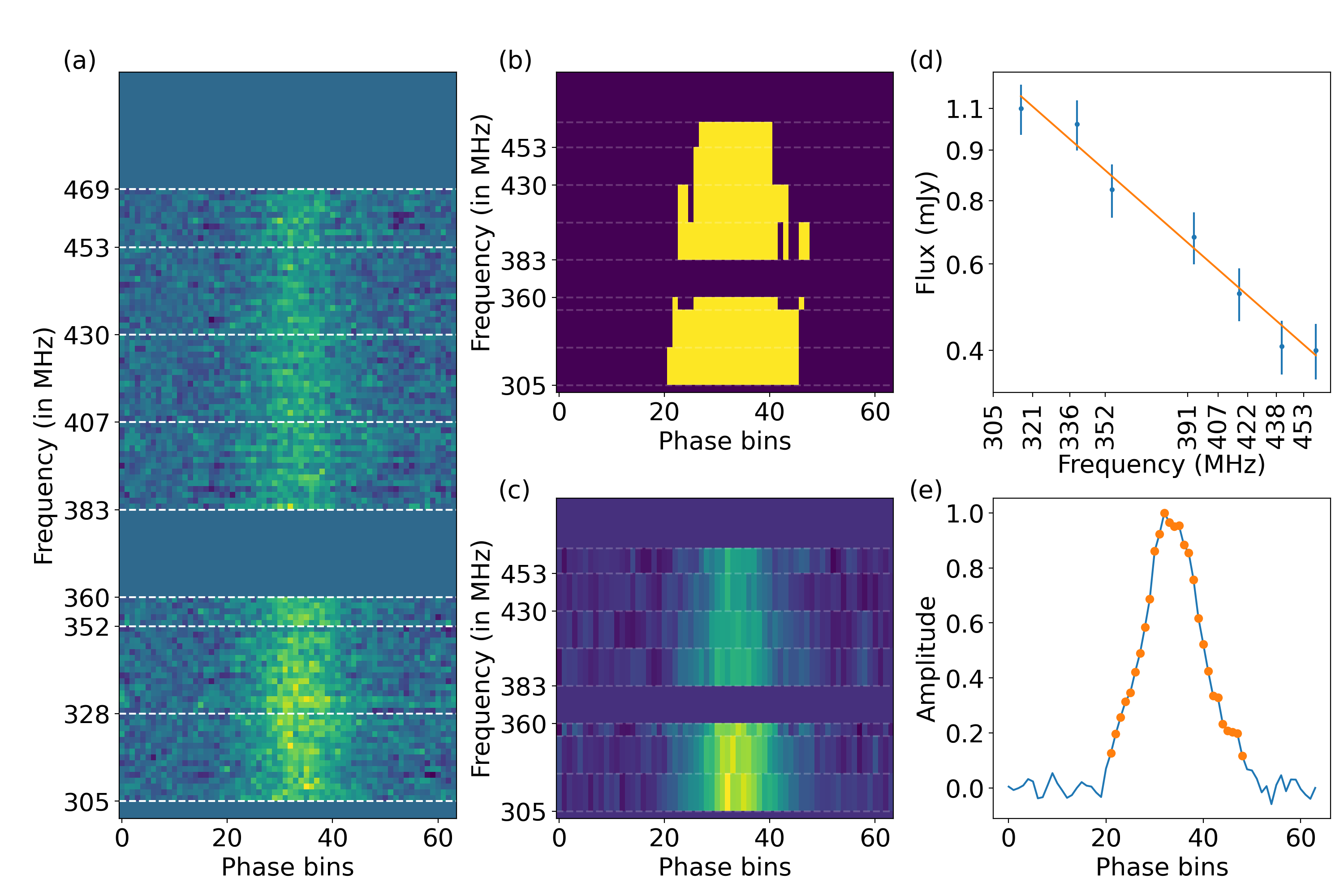}
\hspace*{0.5cm}
\caption{This figure contains various subplots illustrating the functionality of different components of the algorithm for the determination of the spectra for one of the epochs of MSP J1120$-$3618. Panel (a) shows the time-averaged signal (the dashed line shows the edge of the frequency subbands), (b) displays the binary mask of ON phase bins (yellow pixels) and OFF phase bins (violet pixels), (c) the frequency subband averaged flux density, (d) shows the flux density versus frequency, and the orange line shows the fit of the plot, and lastly, plot (e) shows the time and frequency averaged profile (the orange dots represent the ON-phase bins) of the epoch.
\label{fig2:pipeline_outcome}}
\end{figure}

The data cube is divided into frequency 7--8 sub channels (also referred to as subbands which are $\sim25$ MHz) and the flux densities are determined for each subband for each epoch by a pipeline developed by us, using the Python module of the \textsc{PSRCHIVE} \citep{PSRCHIVE_ref}. This pipeline is publicly available on Zenodo \citep{Sharan_pipeline}. The radiometer equation is used to calculate the flux density per subband.
\begin{subequations}\label{eq2:Radiometer_Eq}
\begin{align}
S_{\nu} &= SNR_{\nu} \times RMS_{\nu} \\
RMS &= \frac{T_{sys}}{G \sqrt{N_{ant}^2 N_{pol} \Delta \nu \Delta t}} \sqrt{\frac{W}{P-W}} \\
SNR &= \frac{\sum_{i \in on} (x_i - x_{mean})}{\sigma_{off} \times \sqrt{W}}
\end{align}
where, S$_{\nu}$ is the flux density of the subband with $\nu$ being the central frequency, SNR$_{\nu}$ is the signal-to-noise ratio of the subband (\cite{LorimerKramer}), and RMS$_{\nu}$ is the RMS of the subband. T$_{sys}$ comprised of the T$_{receiver} + $T$_{sky}$. $\Delta \nu$, $\Delta$t, N$_{ant}$ and N$_{pol}$ is the bandwidth of the subband, total time of observation, number of antennas used, and number of polarizations. W and P are the width and period of the pulsar in terms of the number of phase bins. In Appendix \ref{app:on_off_algo}, the method for finding ON phase(W) is discussed in greater detail. Another application of the ON--OFF phase search algorithm involves the exploration of eclipsing and non-eclipsing orbital phases along the subint axis, as demonstrated by \cite{Kumari_et_al_2024} (further details can be found in Appendix \ref{app:on_off_algo}). 
\end{subequations}
Finally, on getting the flux density per subbands, we used AICc \citep[Akaike information criterion with correction for small numbers,][]{Liddle_2007} criterion to fit the optimal number of power law, as done in \cite{Jankowski_et_al_2018}. The optimizing function, AICc, is composed of two parts. The first part is the $\chi^2$ and the second part is the cost function which accounts for the penalty associated with a higher number of model parameters used, described by,
\begin{equation}
AICc = \chi^2 + \frac{2kn}{n-k-1}    
\end{equation}
where $k$ is the number of parameters in the model, $n$ is the number of observations. Figure \ref{fig2:pipeline_outcome} shows an example outcome from the above-mentioned pipeline that calculates the spectral nature of the in-band spectra and fits for the optimal number of broken power laws for MSP J1120$-$3618. 

\subsection{Structure function of flux densities}
\label{Sec 4 : structure_function_of_flux}
The flux density as a function of time (hereafter referred as flux density time series throughout the paper) is not uniformly sampled, as evident from Figure \ref{fig3:All_msp_gal_loc_plot}.  Autocorrelation function (ACF) is a commonly used method to find the correlation scales \citep{Gupta_et_al_1994}, and is especially relevant for the stationary time series. However, flux density time series presented in Figure \ref{fig3:All_msp_gal_loc_plot} hints towards non-stationarity, (e.g. J1828$+$0625, J2144$-$5237, shows a trend of decreasing flux density with epochs). Hence, one of the commonly used methods to estimate the refractive scintillation is through calculation of structure function of flux densities \citep[$D(\tau)$, as done by ][]{You_et_al_2007, Kumamoto_et_al_2020, Gitika_et_al_2023}. For a time lag $\tau$, $D(\tau)$  and its error ($\sigma_D(\tau)$) are given by Equations \ref{eq3:Structure_func},  \ref{eq3:Structure_func_error} \citep{You_et_al_2007};

\begin{subequations}
\begin{align}
\begin{split}
D(\tau) &= \frac{1}{\bar S ^2 N_p} \Bigg[ \sum_{i,j}^{N_p} (S_i - S_j)^2 + \sum_{i,j}^{N_p} (e_i^2 + e_j^2) \label{eq3:Structure_func}\\
 & + 2 \sum_{i,j}^{N_p} (e_i - e_j)(S_i - S_j) - 2 \sum_{i,j}^{N_p} e_ie_j \Bigg]
\end{split}\\
\begin{split}
\sigma_D^2(\tau) &= \frac{1}{\bar S ^4 N_p^2} \Bigg[ \sum_{i} N_i^2e_i^2 + 4\sum_{i,j} e_i^2(S_i - S_j)^2 \label{eq3:Structure_func_error}\\
 &+ 4\sum_{i,j}^{N_p} e_i^2e_j^2 \Bigg]
\end{split}
\end{align}
\end{subequations}
where $\bar S$ is the mean flux density of all the observing epochs, $S_i$, and $S_j$ are average flux densities at individual epochs (represented in MJDs) which are $\tau$ days apart and $N_p$ is the number of such pairs for a given $\tau$. The flux density errors corresponding to $S_i$ (and $S_j$) are denoted by $e_i$ (and $e_j$). One of the advantages of using structure function for the estimation of refractive timescales is that any incorrect estimation of the mean or the variance of the full time series of the flux density would not affect the slope and overall nature of the structure function \citep{Stinebring_et_al_1990, Prokhorov_et_al_1975}. 

\subsection{Structure function of stationary and non-stationary processes}
\label{sec:Structure_func_stationary_non_stationary}
The idea of non-stationarity can be conceived by considering multiple observations of flux density for a particular MSP taken within a short time span, say within a day. One can calculate the statistics (e.g. mean and standard deviation) using all observations conducted over that short time span. This represents the statistics of the population of observed flux densities on that day (assuming Gaussian statistics and that the statistics aren't changing within that short time span). Now, the experiment of multiple observations on the short time span is repeated, for a relatively longer time span, say within a few months, and statistics are compared. If the process is stationary, the statistics (like the mean, standard deviation, auto-correlation function, etc) of the experiments wouldn't change. If not then this would fall under the category of the non-stationary process (which means that the interpretation of quantities like ACF and structure function needs to be revisited). 

Structure function, for a given time lag $\tau$, captures fluctuations in the flux density with a timescale less than $\tau$. A flat structure function over a range of timescales represents that the flux density variations are uncorrelated up to those timescales, which is expected for a stationary process. For stationary time series, the structure function is, $D(\tau) = 2\sigma^2[1 - ACF(\tau)]/\bar S^2$ \citep[][check Appendix \ref{app:struct_to_acf} for the derivation]{Stinebring_et_al_1990}, where $ACF(\tau)$ is the auto-correlation function at a lag of $\tau$. For non-stationary processes, statistics would be similar for nearby epochs (i.e. for smaller values of $\tau$). In other words, for lower values of $\tau$ stationarity holds better. If the $\tau$ is large the first two terms in \ref{structure_function_decomposition} start contributing. As the $\tau$ increases further, the structure function reaches its maximum value ($2m_{ind}^2$, as defined in Section \ref{sec:Temp_variation_of_tot_flux}) and saturates \citep[Figure 5 in][]{Stinebring_et_al_1990}. 

As mentioned in Figure \ref{fig2a:model_structure_func}, the structure function consists of noise, structure, and saturation regimes. The behavior of structure function in the noise regime, at lower time lags (of the order of hours or days), can be attributed to uncorrelated noise sources, like, error of the flux density, pulsar's intrinsic flux density variation, and DISS (variation of flux due to diffractive scintillation) and portion of RISS (refractive scintillation). From Equation \ref{structure_function_decomposition} we can see at low $\tau$, all three terms are smallest. At medium lags \citep[time lags in order of days, which is the structure regime as per,][]{Stinebring_et_al_1990}, structure function follows a power law with respect to lag with a power law index nearly equal to unity \citep{Stinebring_et_al_1990}. If the slope (defined by the power law index) is $\sim 5/3$, the refractive scales follow the Kolmogorov turbulent model \citep{Stinebring_et_al_1990, You_et_al_2007}. As the $\tau$ increases (in the structure regime), all three terms start increasing in Equation \ref{structure_function_decomposition}. Lastly at the largest of the lags, structure function reaches its saturation point, and becomes nearly constant at its highest point. The time lag to reach this point is referred to as the refractive timescale. At this timescale, all three terms in Equation \ref{structure_function_decomposition} saturates. The $ ACF$ is nearly zero because the signals are highly uncorrelated at high-time lags and the flux density mean and standard deviation (the first and second term in Equation \ref{structure_function_decomposition}, respectively) do not change anymore, i.e. we have sampled all the variations of the probability distribution function of the non-stationary process. This suggests that the refractive scintillation timescale (which are not necessarily described by the stationary process) is a combination effect of the stationarity timescales (the timescale at which the first two terms of the Equation \ref{structure_function_decomposition} saturates) and correlation timescales (the timescale at which the $ACF(\tau)$ goes to 0).
\section{RESULTS}\label{results}
\subsection{Temporal variation of flux density}
\label{sec:Temp_variation_of_tot_flux}

\begin{figure*}[h]
\centering
\includegraphics[height=0.6\textheight,width=1\textwidth]{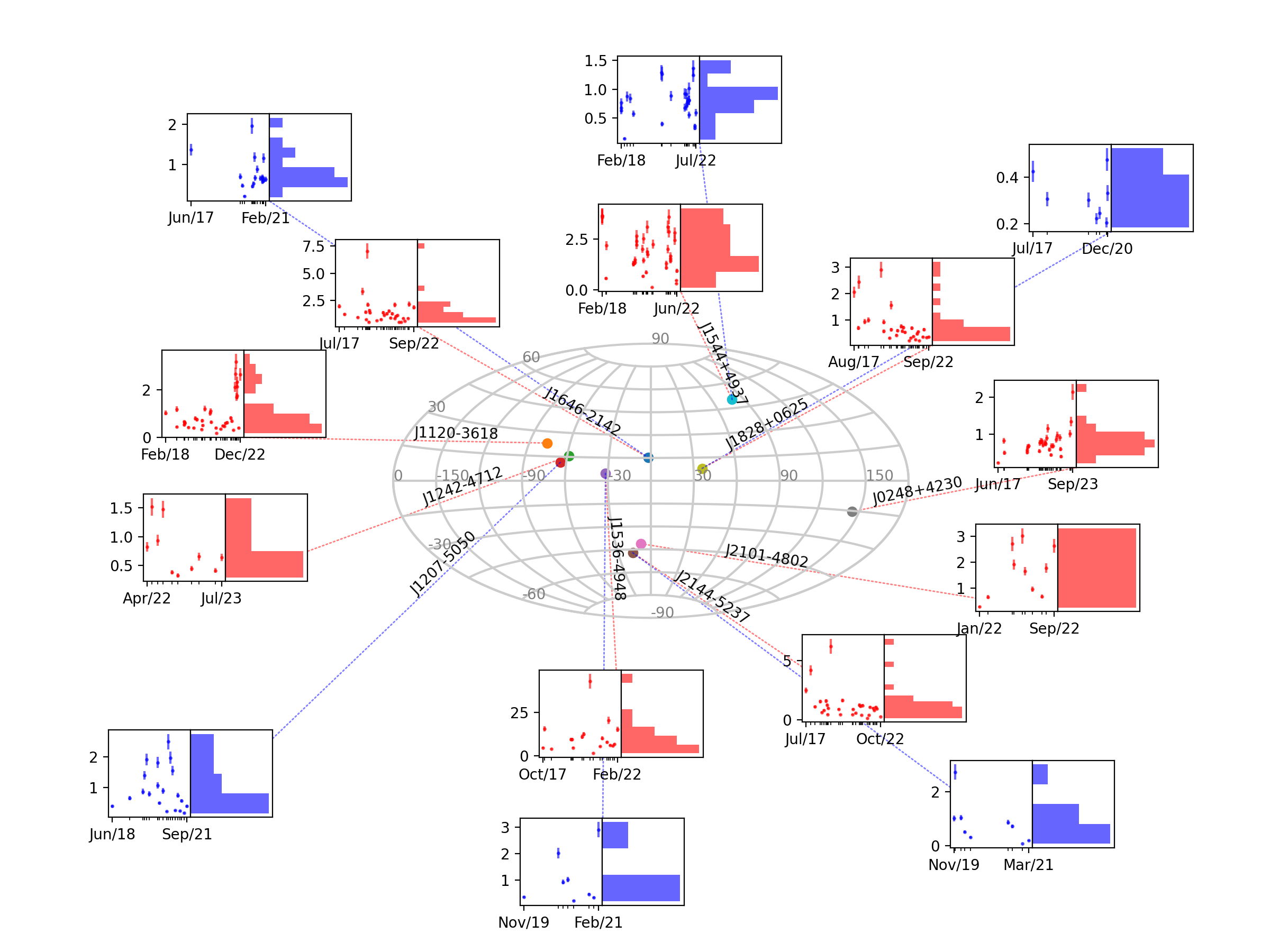}
\caption{Plot for visualization of flux density modulation for the target MSPs. For each MSP, the flux density time series (flux density, with the error bar on flux density, versus the epoch of observation) and its histogram for band 3 (in red) and band 4 (in blue) are presented at their corresponding galactic coordinates.}
\label{fig3:All_msp_gal_loc_plot}
\end{figure*}

\begin{figure}[!hbt]
\centering
    \includegraphics[width=0.45\textwidth]{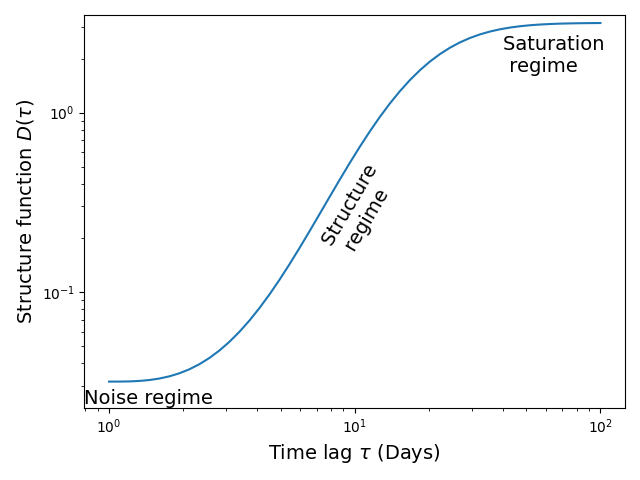}
 
\caption{Plot of the model structure function in log-log scale (similar to the one used in \cite{Stinebring_et_al_1990}).
\label{fig2a:model_structure_func}}
\end{figure}

Figure \ref{fig3:All_msp_gal_loc_plot} shows the location in galactic coordinates and corresponding band averaged flux density time series and histogram for band 3 and 4 for all of the target MSPs. We determined the average flux density, flux density modulation indices, parameter $R$, and average spectral index for bands 3 and 4 (presented in Table \ref{Table3: PSR_summary_info}) across all observing epochs for all the target MSPs. The flux density modulation index is defined by,
\[ m_{ind} = \frac{ \sigma_{S}}{S_{mean}},\] 
where $S_{mean}$ and $\sigma_{S}$ are the mean and standard deviation of flux density across all the epochs of the target MSPs. 
The highest value of the modulation index is observed for J2144$-$5237 amongst all the MSPs, which is 1 in band 3 and 0.9 for band 4. 
Table \ref{Table3: PSR_summary_info} also demonstrates higher modulation indices at lower frequencies for all the MSPs, which is also noted by \cite{Gitika_et_al_2023}. 

The parameter $R$ which is the ratio of maximum to median flux density, presented in Table \ref{Table3: PSR_summary_info}, ranges from 6.4 to 1.7 in band 3 and 4.4 to 1.7 in band 4. We found that the parameter $R$ is correlated with the modulation index in both bands, 3 and 4. Parameter $R$ is more robust than the modulation index \citep[according to][]{Gitika_et_al_2023} as it is less affected by outlier than $m_{ind}$. Hence it can qualitatively suggest the detectability of an MSP due to refractive scintillation. The higher the value of $R$, the higher the flux modulation, and the lower the chance of the MSP being detected, thus making the detection harder. 

Based on the histograms, the majority of pulsars exhibit an exponential tail because our sample consists of low DM MSPs. In non-uniformly sampled data, it is difficult to quantify such trends through ACF. Hence, the structure function was used to calculate the refractive timescale and the model is fitted as described in Section \ref{sec:Model_structure_func}. 

\subsection{Refractive scintillation parameters}
\label{sec:Model_structure_func}
Previous studies on refractive scintillation \citep{Stinebring_et_al_1990, Rickett_et_al_2000} have shown that temporal flux density variation can be best represented by structure function fitted by straight lines on the logarithmic scale. Additionally, an exponential function was suggested for the ACF and hence for the structure function \citep[equation B5 of][]{You_et_al_2007} in the linear scale. The model of structure function we employed to fit the data is inspired by the one presented in \cite{You_et_al_2007} but in the logarithmic scale, as is given below,

\begin{equation}
	\text{log}_{10}D(\tau) = 2(1 - e^{-(\text{log}_{10}\tau/\text{log}_{10}\tau_r)^{\alpha_0}}) + \text{log}_{10}D_{dc}.
\end{equation}

Here, $D(\tau)$ represents the structure function at a time lag $\tau$, $\tau_r$ denotes the refractive timescale and is defined as the point where the exponential term reaches $e^{-1}$, and $D_{dc}$ indicates the structure function value in the noise regime (i.e., at low time lags). The parameter $D_{dc}$ shifts the entire structure function while preserving its features.

\begin{figure*}[h]
\centering
    \includegraphics[width=0.5\textwidth]{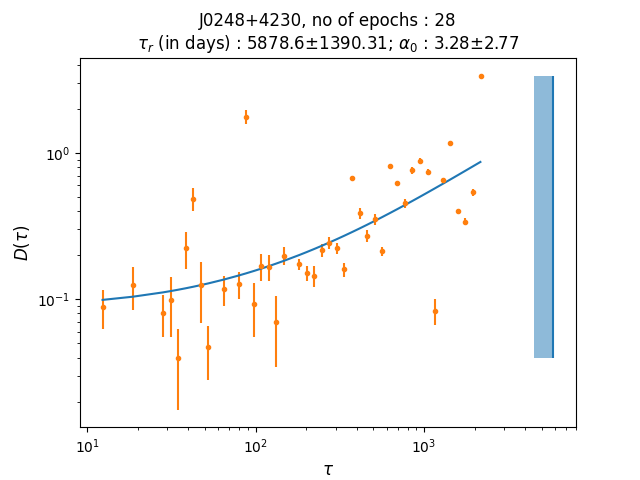}\hfil
    \includegraphics[width=0.5\textwidth]{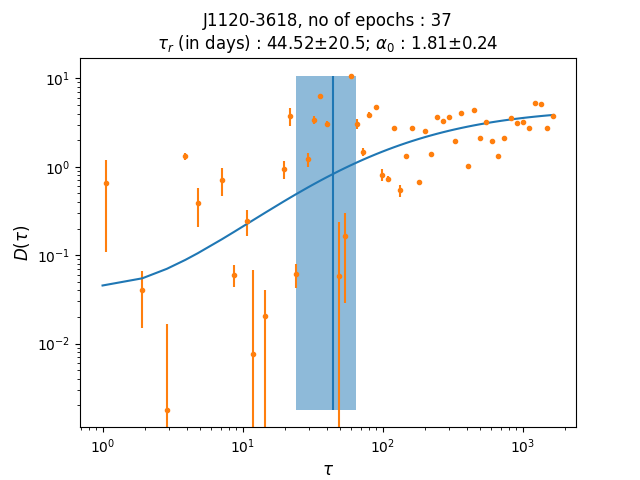}\par\medskip
    \includegraphics[width=0.5\textwidth]{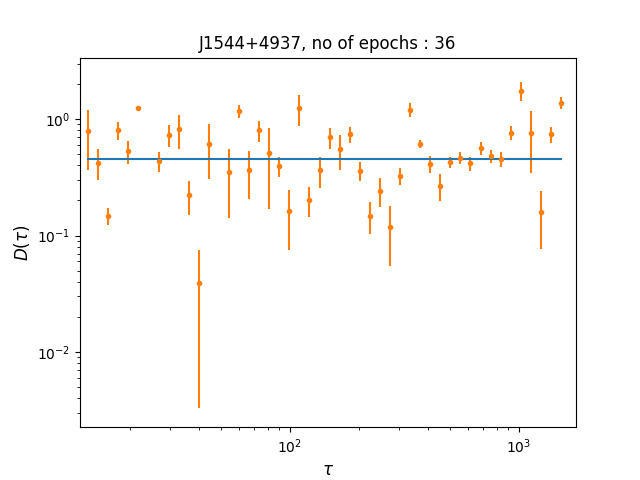}\hfil
    \includegraphics[width=0.5\textwidth]{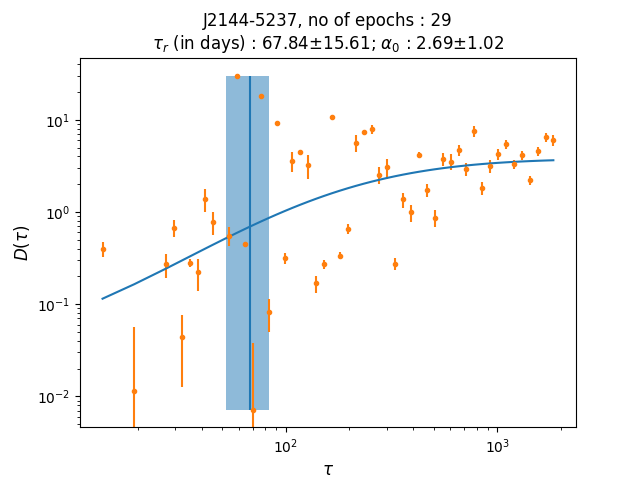}\par\medskip
    \includegraphics[width=0.5\textwidth]{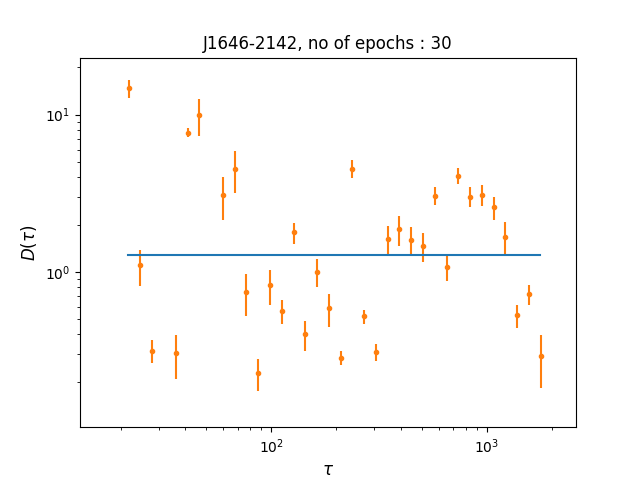}\hfill
    \includegraphics[width=0.5\textwidth]{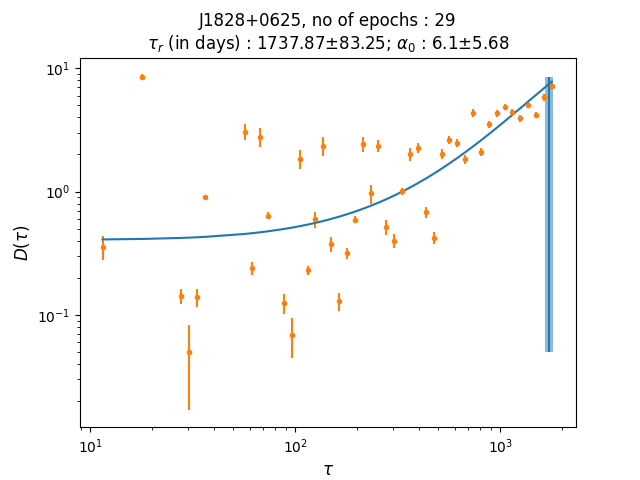}
\caption{Structure function versus the time lag plotted for the target MSPs in band 3.  The respective fits to structure function ($D(\tau)$) using the fitting function $\text{log}_{10}D(\tau) = 2(1 - e^{-(\text{log}_{10}\tau/\text{log}_{10}\tau_r)^{\alpha_0}}) + \text{log}_{10}D_{dc}$(same as equation B5 in \cite{You_et_al_2007}) are shown with blue line. Here, $\tau_r$ (refractive timescale), $\alpha_0$ (the slope of $D(\tau)$ and $\tau$) and $D_{dc}$ (uncorrelated noise dominated regime at small $\tau$ ($\leq$ 1 day)) are estimated by the least square fitting method. The chosen function behaves as plotted in Figure \ref{fig2a:model_structure_func}. The vertical line is the estimated $\tau_r$ and the shaded region is the error of $\tau_r$ (detailed in the Appendix \ref{app: Error_prop_between_linear_and_log_scale}).}
\label{fig4:Structure_function_fit}
\end{figure*}

Figure \ref{fig4:Structure_function_fit} depicts the observed structure function and the corresponding fit to the above described structure function model. Across most of the MSPs examined, the fit of their structure function to the model function indicates a saturation occurring around $\tau \leq 100$ days, except for J1828$+$0625 and J0248$+$4238 where we observe the rise in structure function but did not reach till the saturation regime. The saturation level of structure function from Figure \ref{fig4:Structure_function_fit} is similar to the expected value, which is $2m_{ind}^2$ (as detailed in Section \ref{sec:Structure_func_stationary_non_stationary}). Additionally, the structure function for J1544$-$4937 and J1646$-$2142 lies in the noise regime and does not enter into the structure regime (as explained in Section \ref{sec:Structure_func_stationary_non_stationary}). 

\subsection{Temporal variation of in-band spectra}
\label{subsec:Temp_variation_of_in_band_spectra}
\begin{figure*}[ht]
\centering
\includegraphics[width=0.9\textwidth]{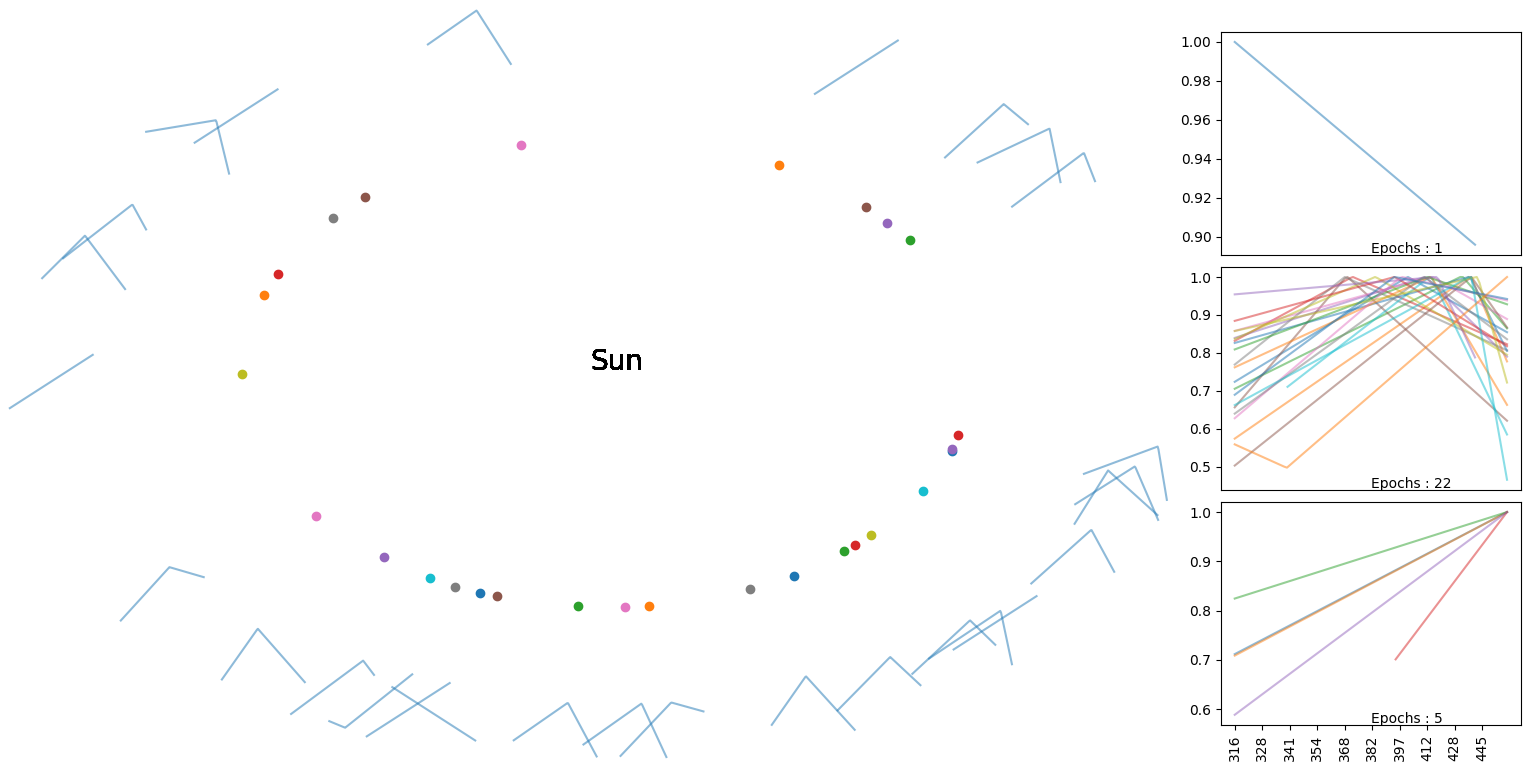}
\caption{The left panel of this plot shows the epoch of observations (represented as different colour filled circles) corresponding to the position of the Earth with respect to the Sun (which is at the center of the elliptical orbit) and their corresponding spectral nature (plotted in the outer ellipse) for J1646$-$2142 observed over 7 years across 30 epochs. On the right panel of the plot, three categories are delineated based on spectral behavior, specifically, spectra with positive indices(top), turnovers(middle), and negative spectral indices(bottom), where x$-$axis is the frequency of the subbands and y$-$axis is the normalized flux density.}
\label{fig5:Temporal_spectral_index}
\end{figure*}

\begin{figure*}
\begin{tabular}{cccc}
\centering
    \includegraphics[height=\Height\textheight,width=\Width\textwidth]{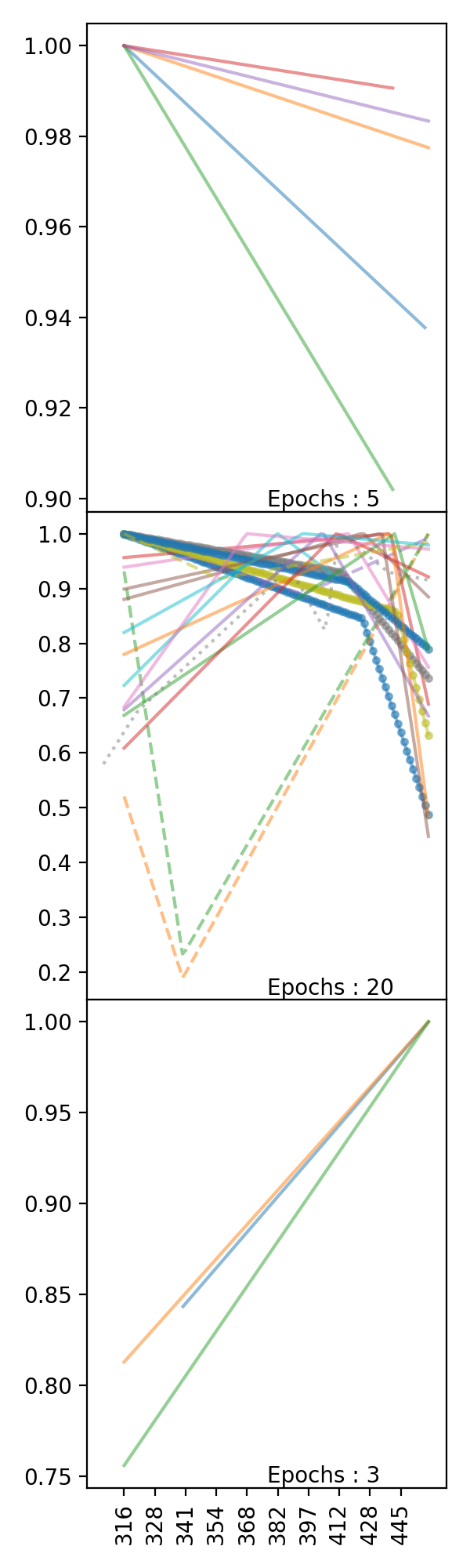} &
    \includegraphics[height=\Height\textheight,width=\Width\textwidth]{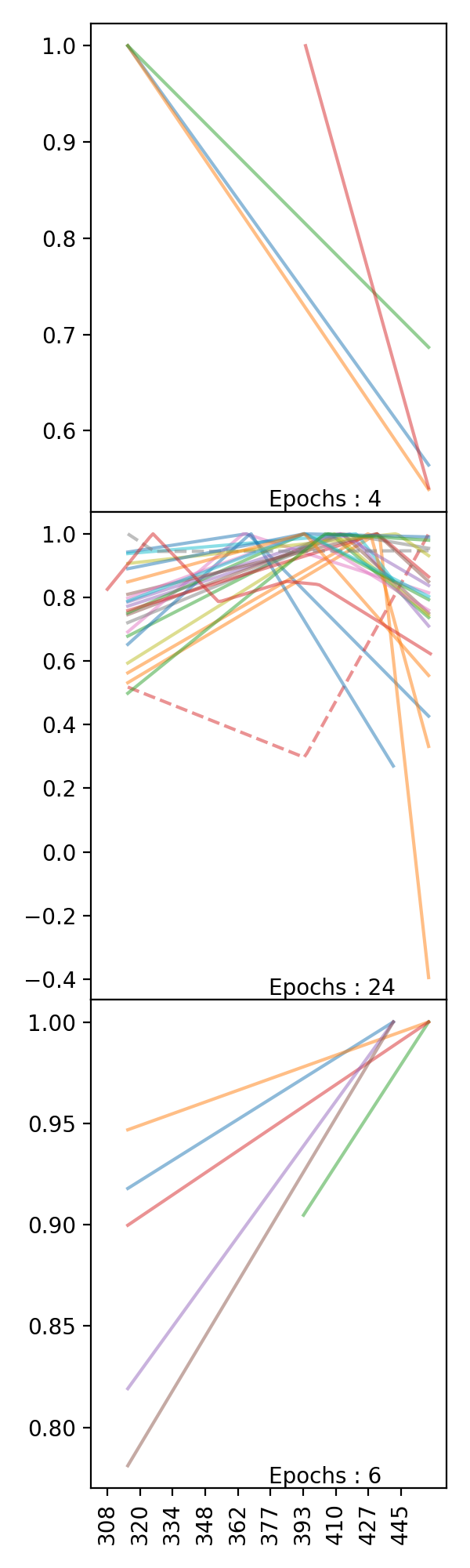} &
    \includegraphics[height=\Height\textheight,width=\Width\textwidth]{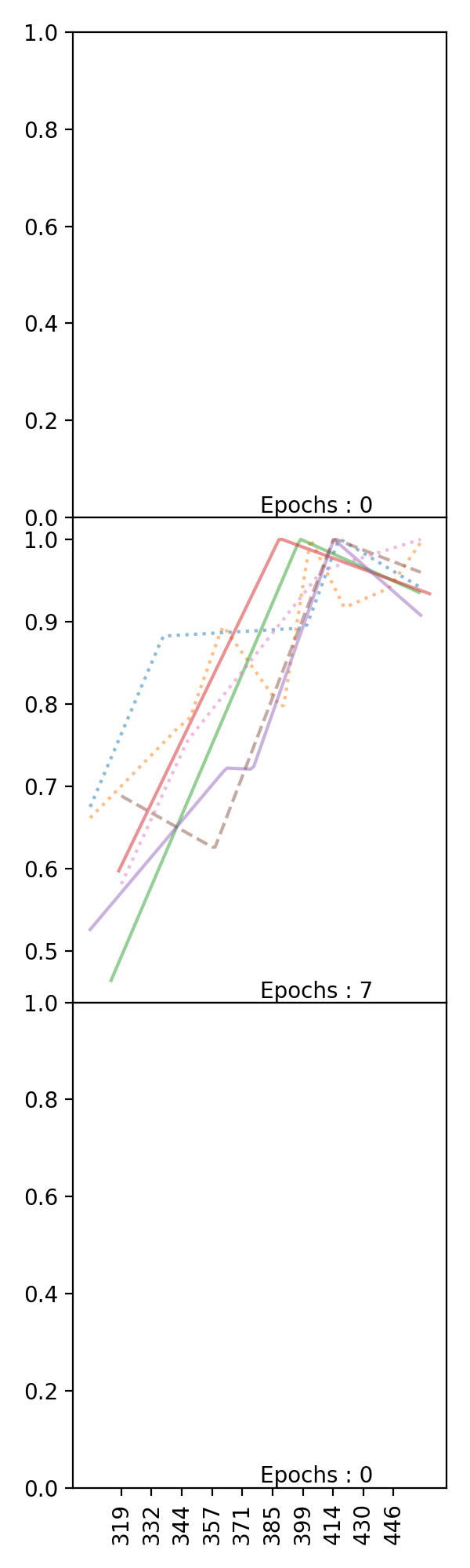} &
    \includegraphics[height=\Height\textheight,width=\Width\textwidth]{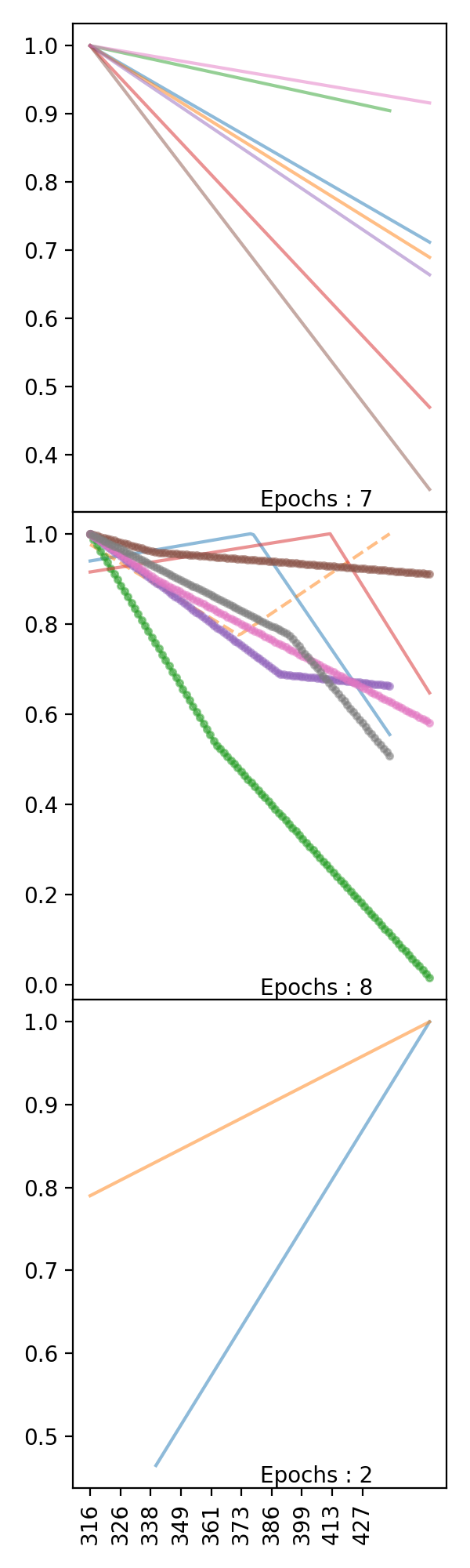} \\
    (a) J0248+4230 (28) & (b) J1120-3618 (37) & (c) J1242-4712 (10) & (d) J1536-4948 (18)\\[5pt]
    \includegraphics[height=\Height\textheight,width=\Width\textwidth]{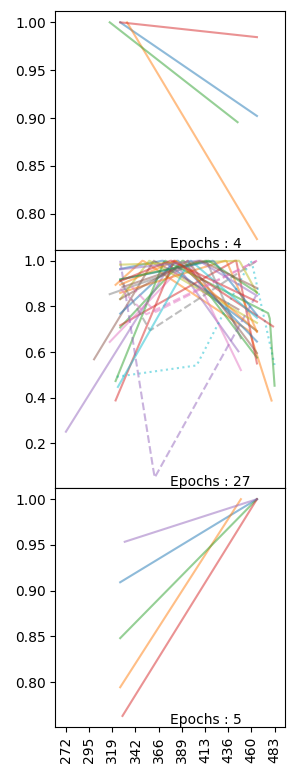} &
    \includegraphics[height=\Height\textheight,width=\Width\textwidth]{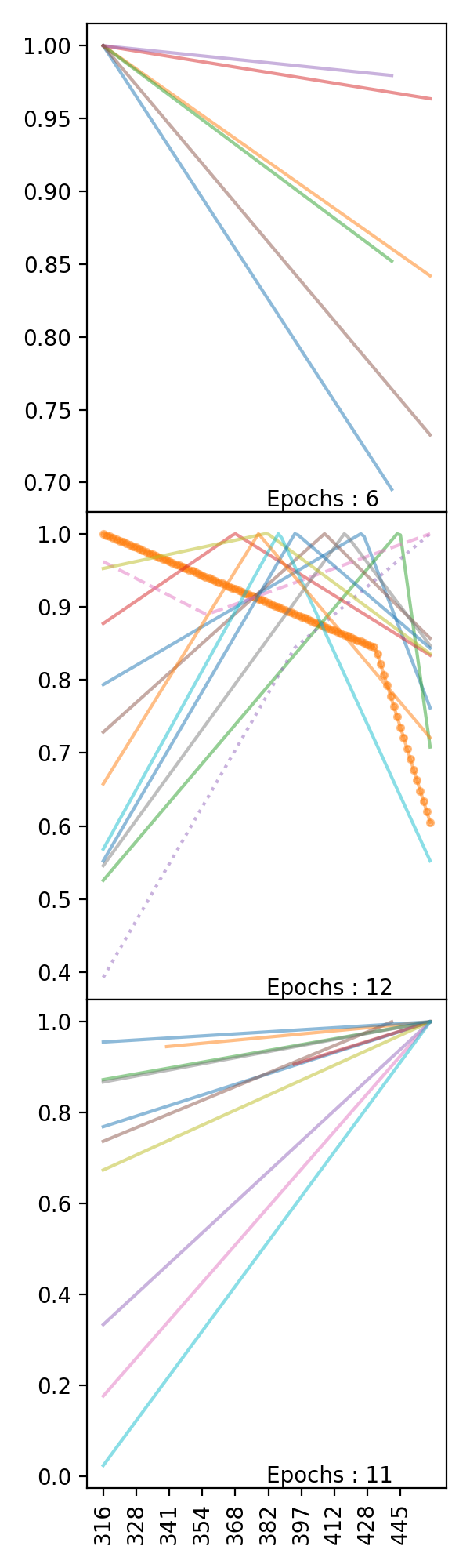} &
    \includegraphics[height=\Height\textheight,width=\Width\textwidth]{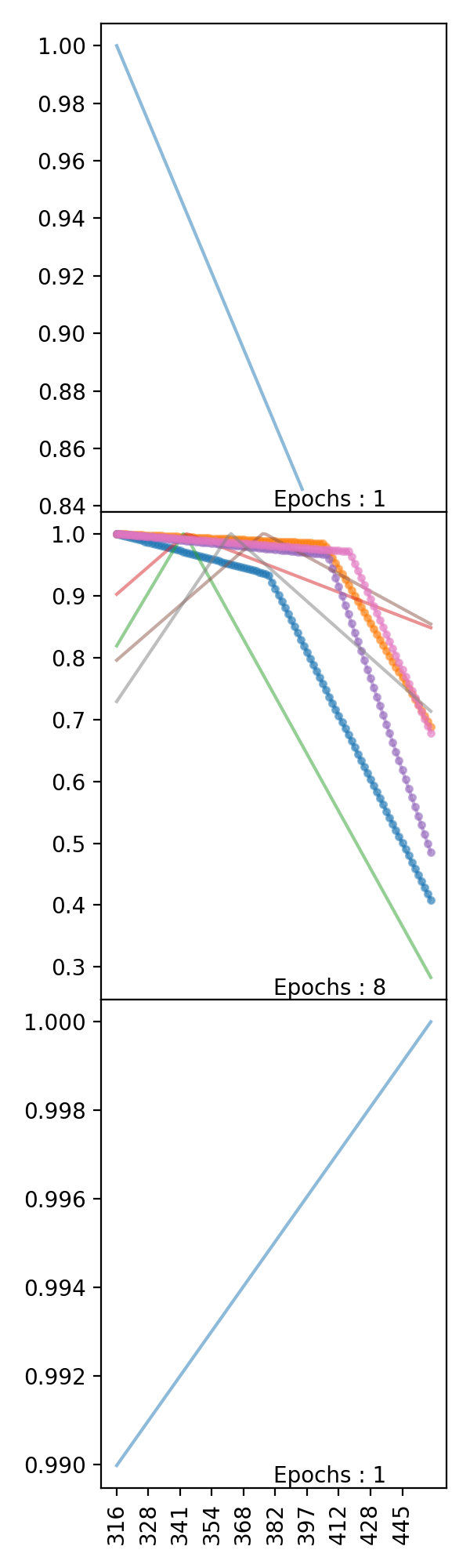} &
    \includegraphics[height=\Height\textheight,width=\Width\textwidth]{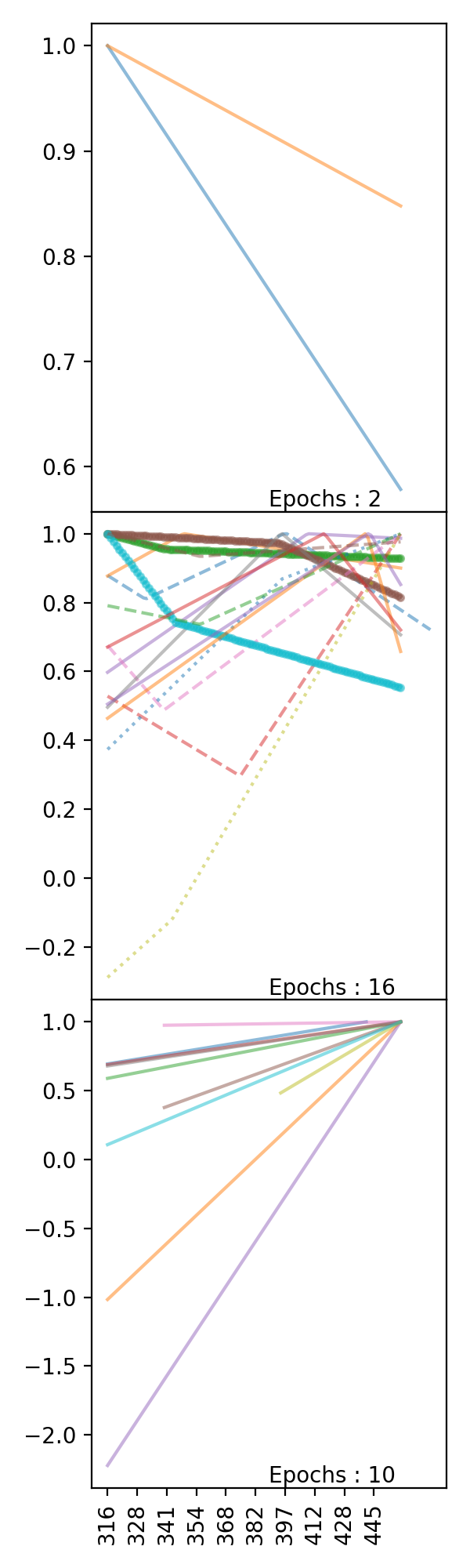} \\
    (e) J1544+4937 (36) & (f) J1828+0625 (29) & (g) J2101-4802 (10) & (h) J2144-5237 (29) \\[5pt]
\end{tabular}
\caption{Variety of spectral nature in band 3 observations of target MSPs (numbers in bracket beside the MSP names represent the total number of epochs, where MSPs having a minimum of 8 observing epochs are considered). The spectra are classified into three categories based on the nature of the best-fit power law function, i.e. single positive spectral indices (top panel), multiple broken power law (middle panel), and single negative spectral indices (bottom panel). The x$-$axis is the frequency and the y$-$axis is the normalized flux density. The flux densities have been normalized by the highest flux density, to aid the visualization of a variety of spectral natures. Different line-styles represent the nature of turnover in the middle plots for all the MSPs. The solid lines represent TO1 type turnover, the dashed line-style represents cases of transition from negative to positive spectral indices, the dotted line-style represents the transition from positive to positive spectral indices, and the connected dots line-style shows the transition from negative to negative spectral indices.}
\label{fig6:all_psr_spectra_categories_b3}
\end{figure*}

We calculated the in-band spectra for the target MSPs and noted huge temporal variations of in-band spectral indices between the observing epochs.
To quantify spectral nature we used the pipeline (described in Section \ref{sec:pipeline_details}) which fits the flux density variation with frequency subbands of the target MSPs using the fitting function comprising of an optimal number of broken power law components (exploiting AICc model selection method, as mentioned in Section \ref{sec:pipeline_details}). Across all the target MSPs, we observed spectral indices ranging from $-6$ to $+6$ in the spectra of individual epochs. The left panel of Figure \ref{fig5:Temporal_spectral_index}, shows the epochs of observation relative to the Sun (at the center of the ellipse). The primary purpose of this plot was to examine any possible correlation between the position of the Earth in the solar system at the epoch of observation and the spectral characteristics of the MSPs, which would be an observational test for the dependence of in-band spectra on the angular distance between the Sun and the MSPs. There was no such correlation evident for any of the target MSPs. The temporal variation of the spectral nature of band 3 for J1646$-$2142 is evident in the right panel of Figure \ref{fig5:Temporal_spectral_index}, where individual epochs are plotted in different colours. The spectra, for different observing epochs, are classified into three categories based on the nature of the best-fit power-law spectra, which are single positive spectral indices (top panel of Figure \ref{fig5:Temporal_spectral_index}), multiple broken power law (middle panel), and single negative spectral indices (bottom panel). Figure \ref{fig6:all_psr_spectra_categories_b3} plots the collection of spectra with uGMRT band 3 observations for the rest of the MSPs, classified in the manner as mentioned above. For all of the target MSPs, the signature of spectral turnover (i.e. transition of negative to positive spectral index) is evident in band 3 from Figures \ref{fig5:Temporal_spectral_index}, and \ref{fig6:all_psr_spectra_categories_b3}. Such turnovers are observed for other pulsars \citep{Kuzmin_et_al_1978, Izvekova_et_al_1981}, from 50 to 380 MHz, commonly known as low-frequency turnovers (referred to as TO1 in the rest of this paper). Additionally, we examined the temporal evolution of the turnover frequency (TO1), noting that some epochs lacked evidence of turnover, while it was observed in others. Along with TO1 (which is transition from positive to negative spectral index), we also noticed other variants of the best-fit spectral nature, such as transition from negative to positive spectral index (dashed line-style in Figure \ref{fig6:all_psr_spectra_categories_b3}), negative to negative (dotted line-styles in Figure \ref{fig6:all_psr_spectra_categories_b3} ), positive to positive (connected dot line-style in Figure \ref{fig6:all_psr_spectra_categories_b3}), in cases of multiple power laws. For example, for J1646$-$2142, out of 30 observing epochs of band 3, we noted TO1 in 21 epochs with frequency for TO1 ranging from 360--430 MHz and 1 epoch with negative to positive change in the spectral index. Along with this, we noted the presence of spectral nature without the evidence of turnover on 6 epochs for J1646$-$2142, the spectra are fitted with a single power law (either with negative or positive spectral index).

\begin{figure*}
\begin{tabular}{ccc}
\centering
    \includegraphics[height=\Heighta\textheight,width=\Widtha\textwidth]{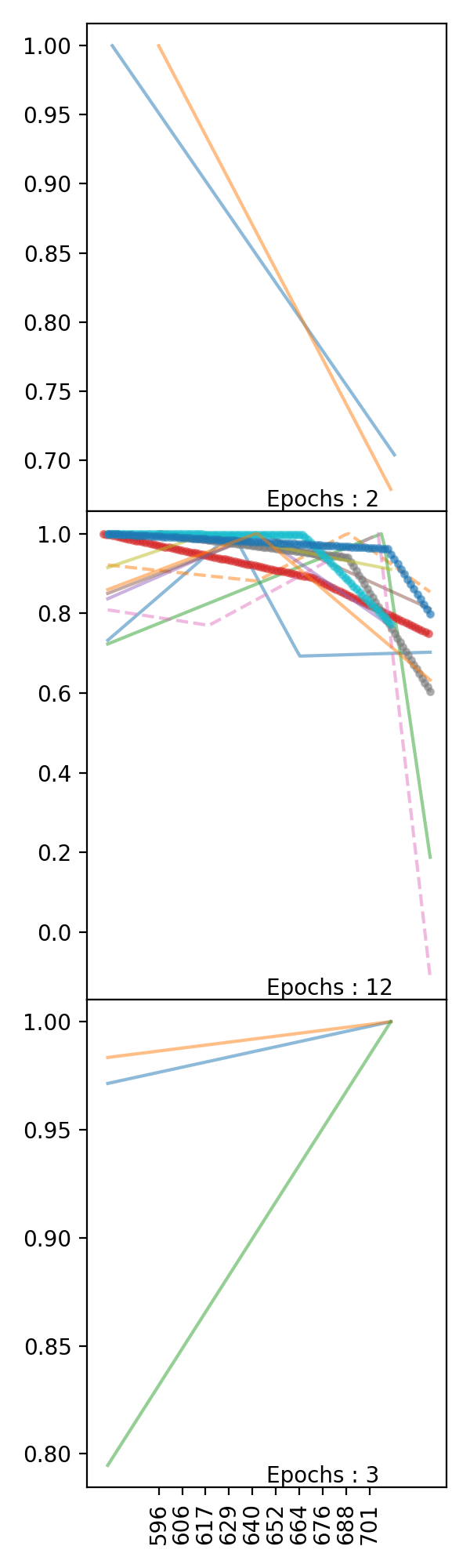} &
    \includegraphics[height=\Heighta\textheight,width=\Widtha\textwidth]{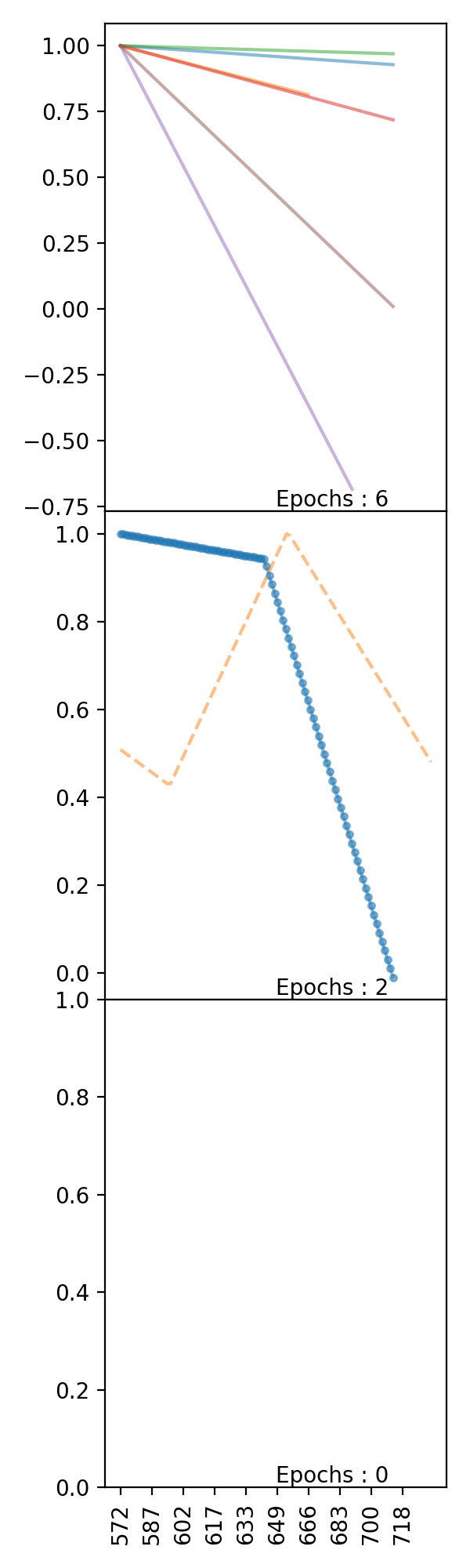} &
    \includegraphics[height=\Heighta\textheight,width=\Widtha\textwidth]{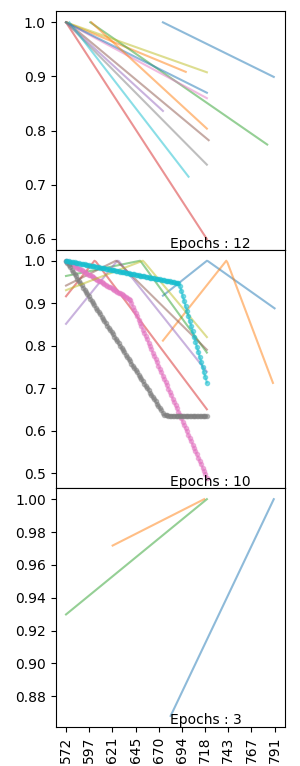} \\
    (a) J1207-5050 (20) & (b) J1536-4948 (8) & (c)J1544+4937 (26) \\[6pt]
    
    \includegraphics[height=\Heighta\textheight,width=\Widtha\textwidth]{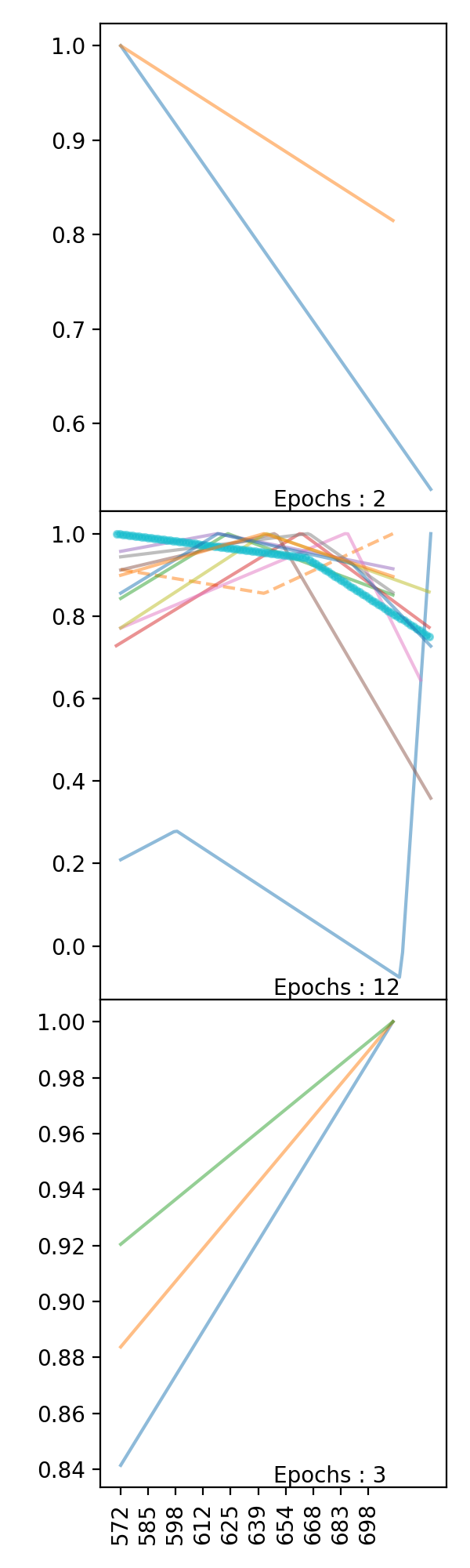} &
    \includegraphics[height=\Heighta\textheight,width=\Widtha\textwidth]{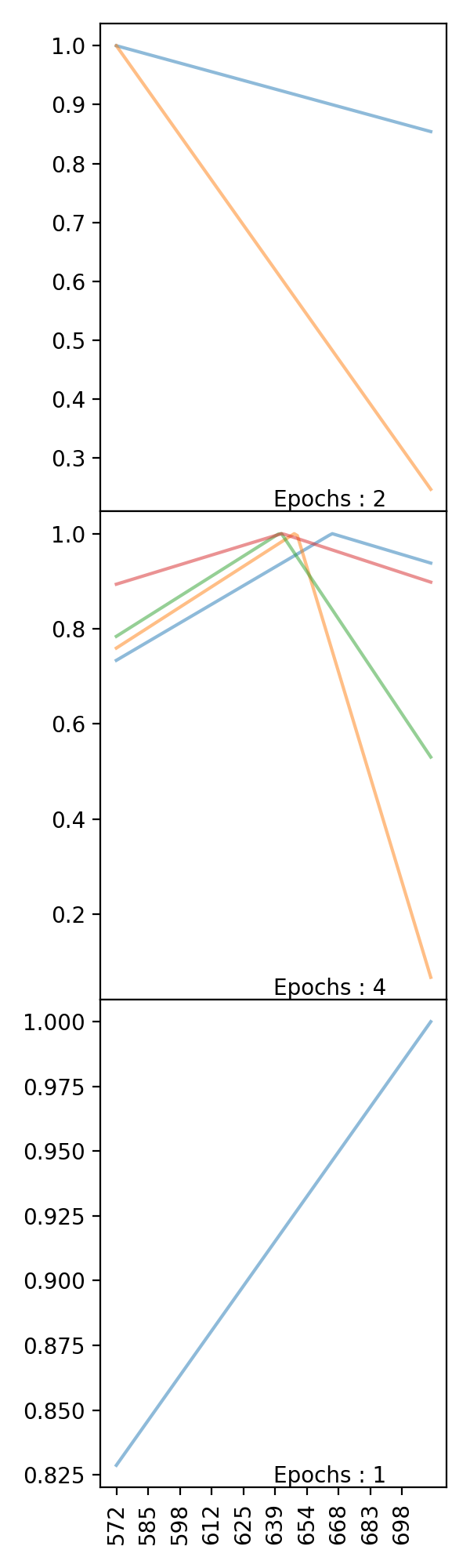} &
    \includegraphics[height=\Heighta\textheight,width=\Widtha\textwidth]{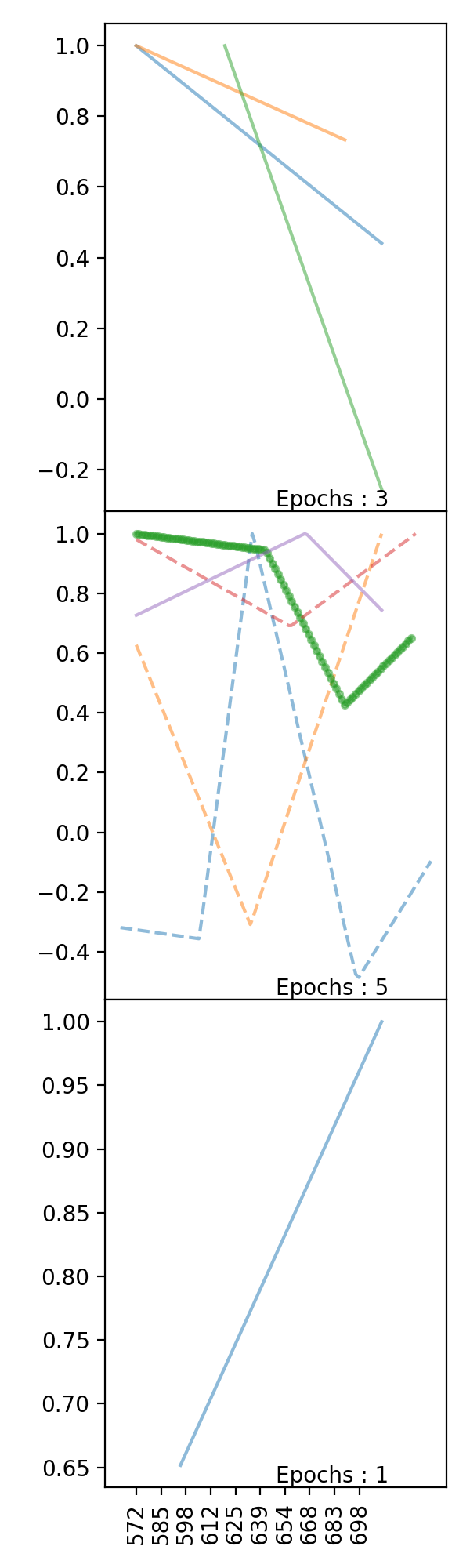} \\
    (d) J1646-2142 (17) & (e) J1828+0625 (8) & (f) J2144-5237 (9) \\[6pt]
\end{tabular}
\caption{Variety of spectral nature in band 4 observations of target MSPs (numbers in bracket beside the MSP name represent the total number of epochs). The description of this Figure remains the same as Figure \ref{fig6:all_psr_spectra_categories_b3}.}
\label{fig7:all_psr_spectra_categories_b4}
\end{figure*}

Figure \ref{fig7:all_psr_spectra_categories_b4} displays spectral nature categories for band 4 data, similar to Figure \ref{fig6:all_psr_spectra_categories_b3}. For example for J1646$-$2142, TO1 frequencies range from 620 to 690 MHz. For almost all target MSPs, we found similar nature of the in-band spectra including the temporal variation of TO1 type turnovers in band 3 and 4.

\section{Discussion}\label{discussion}
This study provides a comprehensive overview of the spectral behavior of 10 MSP discovered using uGMRT, which has not been extensively studied before.
Our estimates of the spectral index($\alpha$), with the full-band and epoch averaged flux density of band 3 and 4, (as mentioned in Table \ref{Table3: PSR_summary_info}), range from $\sim0$ to $-4.8$. Other studies of spectral properties of MSPs using band averaged flux density include, \cite{Foster_et_al_1991} reporting $\alpha$ ranging from $-1.3$ to $-2.6$ using narrow-band receiver of GBT (from 425 MHz to 3 GHz) for 4 MSPs, \cite{Kramer_et_al_1998} reporting the mean $\alpha$ of $-1.8$ observed by narrow-band receivers of Effelsberg radio telescope (centered at 1.3 and 1.7 GHz) for a set of 32 MSPs, \cite{Toscano_et_al_1998} reporting $\alpha$ ranging from $-1.1$ to $-2.9$ in using Parkes narrow-band receiver (436, 660 and 1500 MHz) for 19 MSPs and finally \cite{Dai_et_al_2015} reporting the mean $\alpha$ of $-1.7$ for 24 MSPs. A more recent study by \cite{Gitika_et_al_2023} with the MeerKAT (856--1712 MHz) reported a minimum $\alpha$ of $-3.651$ and a maximum $\alpha$ of $-0.864$ for 89 MSPs. 
Though most of the MSPs from this paper have their $\alpha$ within the range reported by previous studies, we report a relatively higher $\alpha$ of $-4.8$ for MSP J1536$-$4948 respectively.

 We also demonstrated that calculating flux density for subbands in the wide bandwidth is useful when the fractional bandwidth ($x$) is greater than $1$, which is usually the case with either wideband telescopes or telescopes operating at very low frequencies (Section \ref{sec:in-band_spectra}).
The outcome of this long term monitoring of the spectral properties is discussed below.

\subsection{Temporal variation of total flux density}
We observed temporal variation of total flux density as evident from time-series and histogram plots of Figure \ref{fig3:All_msp_gal_loc_plot}. Such variability in flux density can be quantified by modulation index ($m_{ind}$), listed in Table \ref{Table3: PSR_summary_info} for the target MSPs (ranging from 0.99 to 0.45 in band 3 and 0.90 to 0.28 in band 4). An investigation of spectral properties for 89 MSPs by \cite{Gitika_et_al_2023} for the frequency band of 856--1712 MHz reported a modulation index ranging from 2.34 to 0.07. On the other hand, for studies of normal pulsars, \cite{Kumamoto_et_al_2020}, reported modulation index ranges from 1.97 to 0.02 while investigating 286 normal pulsars at 1.4 GHz. 

However, such similarities between observed ranges of modulation index for normal pulsars and MSPs should not be interpreted as caused only by the ISM (e.g. refractive scintillation) acting on the similar intrinsic spectra for the two populations. Breaking the degeneracy amongst the intrinsic variation, the surrounding medium effect and the ISM effect on flux density modulation is not often feasible. Some of the examples of phenomenons where the flux density is affected by intervening medium include scintillation arcs in strong scattering regime \citep{Stinebring_et_al_2001}, plasma lensing \citep{Lin_et_al_2023}, and Gigahertz spectral turnovers \citep{Rajwade_et_al_2015}. 
 
We noted the spider MSPs in our sample have a mean modulation index of $0.53$ (while only considering non-eclipse phase data), whereas non-spider MSPs have a mean modulation index of $0.78$ in band 3. Considering the data from Table 1 of \cite{Gitika_et_al_2023}, we determine the median modulation index of 0.68 for spider MSPs, 0.42 for non spider binary MSPs, and 0.55 for isolated MSPs. There were only 9 spider MSPs, which have mostly low DM, out of 89 MSPs, which could potentially create a bias for a low DM spider MSP sample. 
The reported values of the modulation index may have been influenced by the difference in on-source time for \cite{Gitika_et_al_2023}, which is $<$ 5 minutes, and this study, $\sim$ 45 minutes. To determine if the spider MSPs indeed have a lower modulation index additional data from a larger sample is needed. 

\subsection{Structure-function and its caveats}
Refractive scintillation parameters, such as timescales, and slope of the ACF (which throws light on the nature of the spatial scale of ISM disturbances) have been presented in Table \ref{Table3: PSR_summary_info}. The typical values of the refractive timescales are expected to be in the order of days to months. Some of the target MSPs do not show any rise in the structure function (e.g. J1544$+$4937 and J1646$-$2142 in Figure \ref{fig4:Structure_function_fit}). Investigations by \cite{Wang_et_al_2023, Kumamoto_et_al_2020} also reports similar flat nature of structure function for a handful of pulsars. This could indicate that those MSPs are either in the noise regime or the saturation regime with the total observing spans ($\sim$ 7 years for this study). The latter seems unlikely because this implies that structure function has crossed the noise and structure regime within a day's time lag.

All the above conclusions are based on the stationary process assumption of the flux density (Section \ref{sec:Structure_func_stationary_non_stationary}). The stationary process assumption implies that the statistical properties of ISM, the surrounding environment of the MSPs, or the intrinsic flux density of the MSPs do not change with time. This assumption could be violated in various scenarios like eclipsing systems \citep[e.g.][]{Fruchter_et_Al_1989, Stappers_et_al_1996} and plasma lensing effects \citep[e.g.][]{Lin_et_al_2023}. From the analytical perspective, the ISM imparts its effect through the change in the phase part of the complex electric field vector ($\overrightarrow{E}(k_r, \overrightarrow{p})$, where $k_r$ is the radial direction wave number, $\overrightarrow{p}$ is the sky plane projection of pulsar image, formed during multipath scattering processes by ISM, \cite{Walker_et_al_2004}). In other words, the calculations to quantify the variation of the ISM are evaluated using the normalized electric field \citep{Prokhorov_et_al_1975, Lee_&_Jopikii_1975, Walker_et_al_2004}. Working with only phase terms assumes that the intrinsic pulsar's flux density doesn't change and all the flux density variation is solely due to ISM. 

\subsection{Temporal variation of in-band spectra}
Most available estimates of pulsar flux density (e.g. in the ATNF pulsar catalog\footnote{\url{https://www.atnf.csiro.au/research/pulsar/psrcat/}}) are derived from non-simultaneous measurements, thereby susceptible to both the temporal variability of pulsar intrinsic spectra and the influence of the interstellar medium (ISM).
Only a handful of studies such as \cite{Sieber_1973}, \cite{Kuzmin_et_al_1978}, and \cite{Izvekova_et_al_1981} examined the average flux density variation over different epochs from 0.1--1 GHz frequency range using simultaneous multiple narrow-band observations. 
Nevertheless, utilizing simultaneous observations with narrow-band telescopes will only capture spectral data at two or three points, potentially overlooking variations occurring at a finer frequency scale.
Such flux variation on finer frequency scales would provide insights on absorption features \citep{Rankin_et_al_1983}, turnovers for Gigahertz turnover frequencies \citep[Gigahertz Peaked Spectrum, GPS, ][]{Rajwade_et_al_2015,Kijak_et_al_2011}, scintillation patterns (especially diffractive scintillation pattern having shorter timescale). Other studies, like \cite{Gitika_et_al_2023}, for MSPs, and \cite{Jankowski_et_al_2018, Han_et_al_2016}, mostly for normal pulsars, focused on in-band pulsar spectra by studying the spectral nature of the epoch averaged flux density over the subbands of a wideband telescope, to remove the effects of the only assumed reason for flux density modulation, refractive scintillation.
To our knowledge, none of the existing studies have investigated the temporal fluctuations of flux density.
Thus in addition to averaging the flux density across the epochs (reported in Section \ref{discussion}), we examined the temporal variations in spectral characteristics between multiple observing epochs spanning over 7 years (reported in Section \ref{subsec:Temp_variation_of_in_band_spectra}). We observed huge changes in the spectral index ranging from positive to negative values, and various varieties of turnovers. Such temporal variation of the in-band spectra of MSPs over a timescale of months and years is simplistically attributed to the refractive scintillation caused by the ISM. Refractive scintillation, unlike diffractive scintillation, is known to have broadband frequency variations \citep{Narayan_1992, Rickett_et_al_1984}. Hence such wideband spectral variations are difficult to explain through refractive phenomenon. Low frequency turnover of the spectra (TO1, transition of negative to positive spectral index) are reported by earlier studies\citep{Sieber_1973, Kuzmin_et_al_1978, Izvekova_et_al_1981}. However, to the best of our knowledge the existing studies reporting turnover in the spectral nature for MSPs, have not examined any temporal changes in the TO1.
Furthermore, the presence of TO1 in MSP spectra has been inadequately studied, primarily due to the intrinsic faintness of these objects and the limited availability of wide-bandwidth telescopes until recent times. The only exception is \cite{Wang_et_al_2021}, where the authors indicated presence of a TO1 type turnover with follow-up observations using FAST and Arecibo.
Moreover, we noted signatures of TO1 in some of the epochs in band 4 (Figure \ref{fig7:all_psr_spectra_categories_b4}). Further investigation on such turnover would be useful in breaking the degeneracy of flux density fluctuation from the ISM, surrounding medium, or intrinsic effects, where the in-band spectra would certainly play a role in isolating the broadband from the narrowband phenomena.

\begin{acknowledgments}
We thank the Department of Atomic Energy, Government of India, under project No.12-R\&D-TFR-5.02-0700. The GMRT is run by the National Centre for Radio Astrophysics of the Tata Institute of Fundamental Research, India. We appreciate and acknowledge the support of all the GMRT operators and engineers. We thank Devojyoti Kansabanik, Shyam S. Sharma, and Aswathy Sreekumar for their contribution in acquiring the data.

\end{acknowledgments}

\software{\textsc{PRESTO} \citep{PRESTO_ref},\\
          \textsc{PSRCHIVE} \citep{PSRCHIVE_ref},\\
          \textsc{In-Band Spectra} \citep{Sharan_pipeline}
          }

\bibliography{sample631.bib}{}
\bibliographystyle{aasjournal}

\appendix

\section{Determination of ON-OFF bins and non-eclipsing phase of observation:}
\label{app:on_off_algo}
The data cube has the information of the power of the antennas as a function of time (t), frequency (f), and phase bins (i): \[ I \equiv I(t,f,i)\] For the interest of in-band spectra, in this paper, the profiles were averaged along subband and the time axis. The SNR for each of these profiles was then used to calculate the flux density, as mentioned in Equation \ref{eq2:Radiometer_Eq}. Since Gaussian noise is assumed for I(t,f,i), \[ I_{t,f,i} \in G(\mu_{t,f,i}, \sigma_{t,f,i}) \] where $G(\mu_{t,f,i}, \sigma_{t,f,i})$ is the gaussian distribution with mean $\mu_{t,f,i}$ and sigma $\sigma_{t,f,i}$, which is a function of time, frequency and phase bin. Now if we assume for a given t and f, $\mu_{t,f,i} = \mu_i$ and $\sigma_{t,f,i} = \sigma_i$ (i.e. the noise distribution only changes between ON and OFF phase bin, irrespective of the time and frequency sample), then $\mu_i$ for OFF bins, of the baseline substracted data, would be nearly $0$. Whereas $\mu_i$ for ON bins would be significantly higher than $0$. In that case, the parameter p: \[ p(i) = \frac{\sum_{t,f} I_{t,f,i}}{\sigma_{I_{t,f,i}}\sqrt{t \times f}}\] where $\sigma_{I_{t,f,i}}$ is the standard deviation of the $I$ over t and f, would give a handle on how do we define on/off bins. If the parameter $p(i)$, which is calculated for each phase bin, is greater than the desired threshold (say 3 or 5), then it is labeled as on bin else it is off bin.

The summation is done for each frequency subband and full time of observation, for determining the ON-OFF phase bin. This means the parameter $p(i)$ is sensitive to changes amongst the subbands and any variation in the pulse width within the subband and the observation time is averaged out. The benefit it has over fitting multiple Gaussian profile is that it doesn't care about the number of components to be fitted, and this process can be automated.

To determine the eclipsing and non-eclipsing phases, the summation is done along the frequency of observations, $f$, to estimate the parameter $p(t, i)$, where $t$ is the time of observations divided into several sub-integrations (subints) and $i$ is the phase bin. This creates the two-dimensional binary map (based on the value of $p(t, i)$ greater than or less some threshold) which displays the eclipse subint and ON phase bins.

\section{Algorithm for Structure Function of non uniformly sampled data}

Let the 3 arrays M, S, and E be the MJD array, their corresponding total flux, and its error. To calculate values from the Equations \ref{eq3:Structure_func}, \ref{eq3:Structure_func_error}, we need to probe the pair space of mjd (i.e. we can think of time lag space as pair space of mjd array ) and create a histogram with the value of pair space of S and E. For dealing with the pair space of M, we defined an operation similar to the outer product of a column vector and a row vector but replaced the multiplication with the difference operator, 
\begin{equation*}
m_i \otimes m_j = |m_i - m_j|^2
\end{equation*} , where $m_i$ and $m_j$ are the elements of the array M and timestamp $i$ and $j$ are $\tau$ days apart (where $\tau$ is the time lag in Equation \ref{eq3:Structure_func}). The output of this operator, when applied to the array M, is a square matrix, which now represents the time lag values. Similarly, defining matrix with operations like: square of difference of S, E,etc, other terms of the Equation \ref{eq3:Structure_func}, \ref{eq3:Structure_func_error} are calculated. Now simply taking the histogram on the difference matrix of M and weighing it with appropriate terms calculated above will give the histogram of structure function with respect to time lags.

\subsection{Expansion of Structure function}
\label{app:struct_to_acf}
As per \cite{Tatarskii_book_1971}, the definition of structure function for non-stationary process is :
\begin{align}
\label{app:def_struct_func}
\begin{split}
D(t, \tau) & = \frac{\langle (S(t + \tau) - S(t))^2\rangle}{\bar S^2} \\ 
& = \frac{\langle S^2(t + \tau)\rangle + \langle S^2(t)\rangle -2\langle S(t + \tau) S(t) \rangle}{\bar S^2} \\
& = \frac{1}{\bar S^2}\Bigg[\bar S^2(t + \tau) + \sigma^2(t + \tau) + \bar S^2(t) + \sigma^2(t) \\
& -2\langle S(t + \tau) S(t) \rangle \Bigg]
\end{split}
\end{align}
General form of $ACF(t, \tau)$ (also known as auto-correlation function)is 
\[ ACF(t,\tau) = \frac{\langle S(t + \tau) S(t) \rangle - \bar S(t + \tau) \bar S(t)}{\sigma(t + \tau) \sigma(t)} \]
So
\begin{align} \label{structure_function_decomposition}
\begin{split}
D(t, \tau) & = \frac{1}{\bar S^2} \Bigg[ \bar S^2(t + \tau) + \sigma^2(t + \tau) + \bar S^2(t) + \sigma^2(t) \\
& -2\big(\sigma(t + \tau) \sigma(t) ACF(t, \tau) + \bar S(t + \tau) \bar S(t) \big) \Bigg]\\
& = \frac{1}{\bar S^2} \Bigg[ \bigg(\bar S(t + \tau) - \bar S(t) \bigg)^2 + \sigma^2(t + \tau) + \sigma^2(t) \\
& - 2\sigma(t + \tau) \sigma(t) ACF(t, \tau) \Bigg] \\
& = \frac{1}{\bar S^2} \Bigg[ \bigg(\bar S(t + \tau) - \bar S(t)\bigg)^2 + \bigg(\sigma(t+\tau) -\sigma(t)\bigg)^2 \\
& + 2\sigma(t + \tau)\sigma(t)\bigg(1 - ACF(t, \tau) \bigg)\Bigg]
\end{split}
\end{align}
Since $-1 < ACF(t, \tau) < 1$ all the terms in the square bracket in the equation above are positive. The first two terms in \ref{structure_function_decomposition} depend on stationary timescale, which is the typical timescale $\tau_s$ after which these terms would dominate the structure function and the saturate at the highest point (this is also referred as the saturation regime in the \cite{Stinebring_et_al_1990}). The last term depends on the correlation timescale $\tau_c$. $\tau_0$ from the model fitted in the Figure \ref{fig4:Structure_function_fit} is, thus a combination of both the timescales. Another point to be noted is the fact that since ensemble averaging is usually replaced by the average over time (i.e. the samples are collected over different timestamps are averaged), we assume the process to be ergodic. But since stationarity is necessary condition for an ergodic process, a process can't be ergodic without being stationary. Hence the angular bracket referred in equation \ref{app:def_struct_func}, is the ensemble average and not the time average quantity.

For stationary processes, \cite{Stinebring_et_al_1990}, $\bar S(t + \tau) = \bar S(t) = \bar S $, 
$ \sigma(t + \tau) = \sigma(t) = \sigma_0$ and $ACF$ is independent of $t$, thus from equation \ref{structure_function_decomposition}:
\[ D(\tau) = \frac{2\sigma_0^2\big(1 - ACF(\tau)\big)}{\bar S^2} \]
To put in use case, in the case of stationary processes, even if the mean and standard deviation is not estimated properly, the slope of log$D(\tau)$ vs log$(\tau)$ doesn't get affected.

\section{Error propagation between linear and log scale}
\label{app: Error_prop_between_linear_and_log_scale}
Let $x$ be a linear scale quantity, $\Delta x$ be its error and $X = \text{log}_{10}x$ be the log scale value of $x$, $\Delta X = \Delta \text{log}_{10}x$ be its log scale error of $X$.

\begin{align*}
&X + \Delta X = \text{log}_{10}(x + \Delta x) \\
&\Delta X = \text{log}_{10}(x + \Delta x) - X = \text{log}_{10}(x + \Delta x) - \text{log}_{10}x \\
&\Delta X = \text{log}_{10}\bigg(1 + \frac{\Delta x}{x}\bigg) \Leftrightarrow
\Delta x = x \big(10^{\Delta X} - 1\big) 
\end{align*}

Now when $\frac{\Delta x}{x} < 1$ (this approximation might get violated for errors from the models is taken into consideration, in that case the non-approximated error, mentioned in the above formula, should be applied) then $\Delta X = \frac{\Delta x}{\text{ln}(10)x}$ and $\Delta x = \Delta X \text{ln}(10)x$, where ln refers to log with base e.

\end{document}